# Structural Algebraic Quantum Field Theory*

A. D. Alhaidari

*Saudi Center for Theoretical Physics, P.O. Box 32741, Jeddah 21438, Saudi Arabia*

**Abstract:** Conventional quantum field theory is a method for studying structureless elementary particles. Non-elementary particles, on the other hand, are those with internal structure or particles that are made up of elementary constituents like the hadrons, which contain quarks and gluons. We introduce a structure-inclusive algebraic formulation of quantum field theory that could handle such particles and in which orthogonal polynomials play a central role. For simplicity, we consider non-elementary scalar particles in 3+1 Minkowski space-time and, in three appendices, we treat spinors with structure, massless vector fields, and the massive vector bosons. We show how to do scattering calculation in a nonlinear scalar-spinor coupling model where we find that loop integrals in the Feynman diagrams are remarkably finite. The aim of this short exposé is to motivate further studies and research using this approach.

## 1. Introduction

Conventional quantum field theory (QFT) was developed to describe structureless elementary particles, their interaction with each other and with their environment [1-4]. An example, is the most successful theory that accounts for the electromagnetic interaction of electrons with photons, called quantum electrodynamics (QED) [5-7]. However, in its early days, QFT did not succeed in describing the interaction of nucleons *even at low energies* because they are not elementary. It was later replaced by the more successful quantum chromodynamics (QCD), which is the QFT of quarks and gluons as structureless elementary particles [8,9]. In one of its representations, QFT is visualized by means of diagrams known as the Feynman diagrams that consist of points (vertices) connected by lines (propagators) [10,11]. The lines represent free propagation of elementary particles and the points represent the interaction among particles meeting at those points.

If the particle has a structure† of an infinitesimal size relative to a low energy scale, it is then believed that a faithful QFT could still be used successfully at those low energies. It is only at higher energies that hidden structural effects become significant. Therefore, we adopt the view that a more useful and practical QFT should include particle structure in its formulation since what is believed to be a structureless elementary particle at one energy scale (e.g., the nucleon at KeV energies prior to the 1930s) may turn out to be a composite particle at a higher energy scale. After all, if the particle is indeed structureless for all energies, then that could easily be accommodated in the theory by taking the structure as null. Consequently, we introduce a QFT for particles that may or may not be elementary (i.e., particles that may have internal structures or built from elementary constituents). It will become evident in the ensuing development that the proposed theory has a clear and strong algebraic underpinning. However, it is fundamentally and technically different from that which is commonly known in the mathematics/physics literature as Algebraic Quantum Field Theory (AQFT). A survey of AQFT with an orientation towards foundational topics can be found in [12] and references cited therein. We refer to the theory introduced here as "Structural Algebraic Quantum Field Theory" (SAQFT)[‡] and for simplicity we consider in Section 2 non-elementary scalar particles in 3+1 Minkowski space-time. In Section 3, we present an example of such a particle but in 1+1 space-time. In Section 4, we give an example of a scalar-spinor system with nonlinear coupling and show how to carry out scattering calculations by means of a revised version of the rules of Feynman diagrams appropriate for use in SAQFT. Moreover, in three appendices, we give a brief SAQFT formulation of the Dirac spinor with structure as well as the massive and massless vector fields. The presentation here is elementary and requires basic knowledge in QFT [1-4] and orthogonal polynomials [13-16].

† Particle structure is a finite or infinite set of discrete configurations/states that are locally squeezed within the spatial domain of the particle.

The theory leads to a remarkable numerical result that needs rigorous investigation to see whether it is a general characteristic of SAQFT: In Section 4, we perform scattering calculation in a given scalar-spinor model up to second order in the coupling and find that loop integrals in the Feynman diagrams have finite values eliminating the need for renormalization altogether. It remains to be seen whether this finiteness property is maintained at all loops in the perturbation series and whether it is valid for other physically relevant models as well.

In the published literature, different approaches based on conventional QFT can be found that are rather successful in the description of low energy physics. However, if one strictly applies conventional QFT to describe hadrons at low energies in nuclear physics, then the results obtained would be far off. This is one of the reasons that many articles, books and monographs in nuclear physics introduce and discuss a large number of methods and techniques that are distinct from QFT to handle nucleons and their interactions. The premise here is that SAQFT could be used successfully to treat hadrons as particles whose structure is made up of the constituent quarks without the need for color charges or nonabelian quantum fields. The objective of this brief introductory study is to provide motivation for further investigations and advanced research using this structural algebraic approach to QFT. It is likely to be of interest to physicists working in a variety of fields, including particle physics, nuclear physics, and condensed matter physics. Overall, SAQFT is a promising new approach to QFT with the potential to make significant advances in the field. However, more work is needed to fully develop and apply the approach.

# 2. Scalar particles in SAQFT

In QFT, particle resonances are indispensable objects and of prime physical significance. These objects decay in time (e.g., the free Neutron) whereas stable particles (e.g., the electron) do not. Now, the temporal development of a physical system is governed by the time evolution operator $e^{-\mathrm{i}tH/\hbar}$ with $H$ being the system Hamiltonian, which is a measure of its energy $E$ (i.e., $\left|\psi_t\right\rangle = e^{-\mathrm{i}tH/\hbar}\left|\psi_0\right\rangle = e^{-\mathrm{i}tE/\hbar}\left|\psi_0\right\rangle$). Thus, a simple and clear scheme to account for the decay or stability of these objects is to assign real energies to stable particles whereas resonances are given complex energies with negative imaginary part. Therefore, quantum fields representing particles and resonances could be written/analyzed effectively in the complex energy plane ($E$-plane) or in the complex momentum plane ($k$-plane) since $E$ and $k$ are related by the free field wave equation. Now, in the nonrelativistic theory, the energy-momentum relation for free fields is $E = k^2/2M$ , where $M$ is the rest mass of the particle. Thus, the $k$-plane is more fundamental and unique compared to the $E$-plane since a single point in the $k$-plane corresponds to two points in the $E$-plane that are located on two separate complex "energy sheets." Hence, one $E$-plane is not physically unique while one $k$-plane is. However, in the relativistic theory and using the relativistic units, $\hbar = c = 1$, the energy-momentum relation for free fields is $E^2 = k^2 + M^2$. Thus, the two complex planes are physically equivalent and in SAQFT we choose to work in the $E$-plane rather than the $k$-plane. However, the physical information about the three-space directional degrees of freedom carried by $\vec{k}$ in the conventional formulation must be preserved in the new formulation. Consequently, the zero-spin Klein-Gordon quantum field in 3+1 Minkowski space-time is represented in SAQFT by the following Fourier expansion in the *energy*

$$\Psi(t,\vec{r}) = \int_{\Omega} e^{-iEt}\psi(E,\vec{r})\,a(E)\,dE + \sum_{j=0}^{N} e^{-iE_j t}\psi_j(\vec{r})\,a_j \,. \tag{1}$$

[‡] An easy pronunciation of SAQFT would be "sak FT"

The integral represents the continuous energy spectrum of the particle whereas the sum represents its discrete spectrum (i.e., the particle structure resolved in the energy domain). The latter is a new addition to conventional QFT that could have a positive impact on the treatment of complex systems. The structure in (1) consists of $N+1$ discrete channels whereas $\Omega$ consists generally of several disconnected but continuous channels (called energy bands or energy intervals). These channels do not overlap (i.e., their intersection is null). For QFT applications in particle physics, we take $\Omega$ to stand for the single semi-infinite energy interval $E^2 \geq M^2$ whereas $0 \leq E_j^2 < M^2$. A direct and straightforward implication of such an energy spectrum designation is that massless particles (e.g., the photon) in SAQFT are automatically structureless. The objects $a(E)$ and $a_j$ are field operators (the vacuum annihilation operators). They satisfy the following conventional commutation relations [1-4]

$$\left[a(E),a^\dagger(E')\right] := a(E)a^\dagger(E') - a^\dagger(E')a(E) = \delta(E-E'), \quad \left[a_i,a_j^\dagger\right] = \lambda^{-1}\delta_{i,j}, \tag{2}$$

where $\lambda$ is a real positive scale parameter of inverse length dimension that gives a measure of the structure size and/or mass.[§] All other commutators among $a(E)$, $a^\dagger(E)$, $a_j$, and $a_j^\dagger$ vanish.

The continuous Fourier energy component $\psi(E,\vec{r})$ in (1) has a spatial range that extends all the way to infinity whereas the discrete component $\psi_j(\vec{r})$ has a short range and is locally clustered within the particle spatial domain. These are written as the following pointwise convergent series

$$\psi(E,\vec{r}) = \sum_{n=0}^{\infty} f_n(E)\phi_n(\vec{r}) = f_0(E)\sum_{n=0}^{\infty} p_n(z)\phi_n(\vec{r}), \tag{3a}$$

---

[§] One may consider an alternative definition of the commutation relation for the discrete creation/annihilation operators as $\left[a_i,a_j^\dagger\right] = \delta_{i,j}/(E_j+M)$. In this case the formula $g_0^2(E_j) = \lambda\xi(z_j)$ written below Eq. (11) must be changed to read $g_0^2(E_j) = (E_j+M)\xi(z_j)$. Nonetheless, other formulas containing the scale parameter $\lambda$ like Eq. (8) and Eq. (14) should not be altered.

$$\psi_j(\vec{r}) = \sum_{n=0}^{\infty} g_n(E_j)\phi_n(\vec{r}) = g_0(E_j)\sum_{n=0}^{\infty} p_n(z_j)\phi_n(\vec{r}) . \quad (3b)$$

where $z$ is a real energy parameter (we call it the *spectral parameter*) to be determined, and $\{f_n, g_n\}$ are *real* expansion coefficients which are written as $f_n = f_0 p_n$ and $g_n = g_0 p_n$ making $p_0 = 1$. $\{\phi_n(\vec{r})\}$ is a complete set of functions in configuration space that satisfy the following differential relation

$$-\vec{\nabla}^2\phi_n(\vec{r}) = \alpha_n\phi_n(\vec{r}) + \beta_{n-1}\phi_{n-1}(\vec{r}) + \beta_n\phi_{n+1}(\vec{r}) , \quad (4)$$

where $\vec{\nabla}^2$ is the three-dimensional Laplacian and $\{\alpha_n, \beta_n\}$ are real constants that are evidently independent of the energy and $z$ and such that $\beta_n \neq 0$ for all $n$. Using Eq. (4) in the free Klein-Gordon wave equation $\left(\partial_t^2 - \vec{\nabla}^2 + M^2\right)\Psi(t,\vec{r}) = 0$, we obtain the following equivalent algebraic relation

$$z\,p_n(z) = \alpha_n p_n(z) + \beta_{n-1} p_{n-1}(z) + \beta_n p_{n+1}(z) , \quad (5)$$

for $n = 1,2,3,...$ and with $z = E^2 - M^2$, $z_j = E_j^2 - M^2$. This equation along with its consequential and ensuing relations like the orthogonality relation (6) and completeness formulas (8) shown below represent the on-shell conditions for the quantum field (1). Equation (5) is a symmetric three-term recursion relation that makes $\{p_n(z)\}$ a sequence of polynomials in $z$ (we call them the *spectral polynomials*) with the two initial values $p_0(z) = 1$ and $p_1(z)$ linear in $z$. Now, Eq. (5) has two linearly independent polynomial solutions. We choose the solution with the initial values $p_0(z) = 1$ and $p_1(z) = (z - \alpha_0)/\beta_0$. Due to Favard theorem [17,18] (a.k.a. the spectral theorem; see Section 2.5 in [16]), the polynomial solutions of Eq. (5) satisfy the following general orthogonality relation [13-16]

$$\int_{\Omega} \rho(z) p_n(z) p_m(z) dz + \sum_{j=0}^{N} \xi(z_j) p_n(z_j) p_m(z_j) = \delta_{n,m} , \quad (6)$$

where $\rho(z)$ is the continuous component of the weight function and $\xi(z_j)$ is the discrete component. These weight functions are positive definite and will be determined below in terms of $f_0(E)$ and $g_0(E_j)$, respectively. The fundamental algebraic relation (5), which is equivalent to the Klein-Gordon wave equation, is the reason behind the algebraic setup of the theory and for which we qualify this QFT as algebraic. In fact, postulating the three-term recursion relation (5) eliminates the need for specifying a free field wave equation. Furthermore, once the set of spectral polynomials $\{p_n(z)\}$ is given then all physical properties of the corresponding particle are determined. Thus, a physical process in SAQFT is equivalent to calculating the change from the set $\{p_n(z)\}$ to another set $\{q_n(y)\}$ due to this process. For elastic scattering, the identity of the particles is retained and $\{p_n(z)\} \mapsto \{p'_n(z')\}$. However, for inelastic scattering, $\{p_n(z)\} \mapsto$ $\{q_n(y)\}$, where $\{q_n(y)\}$ is a different set of spectral polynomials with its own spectral parameter, recursion relation and orthogonality corresponding to different outgoing particles.

In conventional QFT, the quantum field (1) is expressed as Fourier expansion in the linear momentum $\vec{k}$-space not in the energy space. That is, $\Psi(t,\vec{r})$ is written as the integral $\int e^{-\mathrm{i}Et+\mathrm{i}\vec{k}\cdot\vec{r}} a(\vec{k}) \frac{d^3k}{\sqrt{(2\pi)^3 2E}}$, where $\vec{k}^2 = E^2 - M^2$ and giving $\psi(E,\vec{r}) \propto e^{\mathrm{i}\vec{k}\cdot\vec{r}} \big/ \sqrt{E}$. Nonetheless, one can show that $e^{\mathrm{i}\vec{k}\cdot\vec{r}}$ could be written as an infinite series having the same SAQFT form (3a) by using the relation

$$e^{\mathrm{i}kx} = \sqrt{2}\, e^{-z^2/2} e^{-\lambda^2 x^2/2} \sum_{n=0}^{\infty} \frac{\mathrm{i}^n}{2^n n!} H_n(z) H_n(\lambda x), \tag{7a}$$

where $z = k/\lambda$, $H_n(y)$ is the Hermite polynomial, $\lambda$ is a real scale parameter, and $\phi_n(x) \propto e^{-\lambda^2 x^2/2} \left[ \mathrm{i}^n H_n(\lambda x) \big/ \sqrt{2^n n!} \right]$. Therefore, $p_n(z) = H_n(z) \big/ \sqrt{2^n n!}$ with $\alpha_n = 0$, $\beta_n = \sqrt{(n+1)/2}$, and $f_0(E) = \pi^{-\frac{1}{4}} e^{-z^2/2}$. An equivalent expansion could also be written in terms of products of the Gegenbauer (ultra-spherical) polynomials $\left\{ C_n^\nu(z) \right\}$ and the Bessel function with discrete index $J_{n+\nu}(\lambda x)$ as follows, (see Eq. (4.8.3) in Ref. [16])

$$e^{\mathrm{i}kx} = 2^\nu \Gamma(\nu) \left( \lambda x \right)^{-\nu} \sum_{n=0}^{\infty} \mathrm{i}^n (n+\nu) C_n^\nu(z) J_{n+\nu}(\lambda x), \tag{7b}$$

giving $p_n(z) \propto C_n^\nu(z)$ and $\phi_n(x) \propto \mathrm{i}^n (n+\nu) \left( \lambda x \right)^{-\nu} J_{n+\nu}(\lambda x)$. In the last paragraph of this section (see Table 1), we show that the spatial representation of the quantum field (1) at any given time is, in fact, a generalization of the quantum field representation in conventional QFT.

Now, the conjugate quantum field $\bar{\Psi}(t,\vec{r})$ in SAQFT is obtained from (1) by complex conjugation and the replacement $\phi_n(\vec{r}) \mapsto \bar{\phi}_n(\vec{r})$ where

$$\left\langle \phi_n(\vec{r}) \middle| \bar{\phi}_m(\vec{r}) \right\rangle = \left\langle \bar{\phi}_n(\vec{r}) \middle| \phi_m(\vec{r}) \right\rangle = \frac{1}{\lambda} \delta_{n,m}, \tag{8a}$$

$$\sum_{n=0}^{\infty} \phi_n(\vec{r}) \bar{\phi}_n(\vec{r}\,') = \sum_{n=0}^{\infty} \bar{\phi}_n(\vec{r}) \phi_n(\vec{r}\,') = \frac{1}{\lambda} \delta^3(\vec{r} - \vec{r}\,'). \tag{8b}$$

In $n+1$ space-time, this makes the length scale of $\phi_n(\vec{r})$ and $\bar{\phi}_n(\vec{r})$ equals $\frac{1}{2}(1-n)$. Equation (8a) is the orthogonality relation whereas (8b) is the completeness statement. Therefore, we write $\bar{\Psi}(t,\vec{r})$ as follows

$$\bar{\Psi}(t,\vec{r}) = \int_\Omega e^{\mathrm{i}Et} \bar{\psi}(E,\vec{r}) a^\dagger(E)\, dE + \sum_{j=0}^{N} e^{\mathrm{i}E_j t} \bar{\psi}_j(\vec{r}) a_j^\dagger. \tag{9}$$

where the components $\bar{\psi}(E,\vec{r})$ and $\bar{\psi}_j(\vec{r})$ are identical to (3) but with $\phi_n(\vec{r}) \mapsto \bar{\phi}_n(\vec{r})$. Using the commutators (2) of the field operators $a(E)$ and $a_j$, we can write

$$\begin{aligned} \left[ \Psi(t,\vec{r}), \bar{\Psi}(t',\vec{r}\,') \right] &= \sum_{n,m=0}^{\infty} \phi_n(\vec{r}) \bar{\phi}_m(\vec{r}\,') \\ \times &\left[ \int_\Omega e^{-\mathrm{i}E(t-t')} f_0^2(E) p_n(z) p_m(z)\, dE + \lambda^{-1} \sum_{j=0}^{N} e^{-\mathrm{i}E_j(t-t')} g_0^2(E_j) p_n(z_j) p_m(z_j) \right] \end{aligned} \tag{10}$$

The general orthogonality (6) and the completeness (8b) turn Eq. (10) with $t = t'$ into

$$\left[\Psi(t,\vec{r}),\bar{\Psi}(t,\vec{r}')\right] = \frac{1}{\lambda}\delta^3(\vec{r}-\vec{r}'), \tag{11}$$

provided that we take $f_0^2(E)dE = \rho(z)dz$ and $g_0^2(E_j) = \lambda\xi(z_j)$, which also imply positivity of the two weight functions with $\frac{dz}{dE} > 0$ for $E \in \Omega$. Moreover, it is straightforward to write

$$\left[\Psi(t,\vec{r}),\Psi(t,\vec{r}')\right] = \left[\bar{\Psi}(t,\vec{r}),\bar{\Psi}(t,\vec{r}')\right] = 0. \tag{12}$$

In the canonical quantization of fields, the canonical conjugate to $\Psi(t,\vec{r})$ is written as $\Pi(t,\vec{r})$ and they satisfy the following equal time commutation relations [1-4]

$$\left[\Psi(t,\vec{r}),\Psi(t,\vec{r}')\right] = \left[\Pi(t,\vec{r}),\Pi(t,\vec{r}')\right] = 0, \tag{13a}$$

$$\left[\Psi(t,\vec{r}),\Pi(t,\vec{r}')\right] = \mathrm{i}\delta^3(\vec{r}-\vec{r}'). \tag{13b}$$

Therefore, we obtain the following identification: $\Pi(t,\vec{r}) = \mathrm{i}\lambda\bar{\Psi}(t,\vec{r})$. Moreover, in analogy with conventional QFT [1-4], we can write Eq. (10) as

$$\left[\Psi(t,\vec{r}),\mathrm{i}\lambda\bar{\Psi}(t',\vec{r}')\right] = \mathrm{i}\Delta(t-t',\vec{r}-\vec{r}'), \tag{14}$$

where the singular function $\Delta(t-t',\vec{r}-\vec{r}')$ in SAQFT reads as follows

$$\Delta(t-t',\vec{r}-\vec{r}') =$$
$$\lambda\sum_{n,m=0}^{\infty}\phi_n(\vec{r})\bar{\phi}_m(\vec{r}')\left[\int_\Omega e^{-\mathrm{i}E(t-t')}\rho(z)p_n(z)p_m(z)dz + \sum_{j=0}^{N}e^{-\mathrm{i}E_j(t-t')}\xi(z_j)p_n(z_j)p_m(z_j)\right] \tag{15}$$

Moreover, Eq. (11) and Eq. (14) give: $\Delta(0,\vec{r}-\vec{r}') = \delta^3(\vec{r}-\vec{r}')$.

Now, we can define the real (neutral) scalar particle with structure by the quantum field $\Phi(t,\vec{r}) = \frac{1}{\sqrt{2}}\left[\Psi(t,\vec{r}) + \bar{\Psi}(t,\vec{r})\right]$ with $\bar{\phi}_n(\vec{r}) = \phi_n^\dagger(\vec{r})$. On the other hand, the complex (charged) scalar non-elementary particle is defined by the positive-energy quantum field

$$\Phi(t,\vec{r}) = \frac{1}{\sqrt{2}}\left[\Psi_+(t,\vec{r}) + \Psi_-^\dagger(t,\vec{r})\right]. \tag{16a}$$

$\Psi_\pm(t,\vec{r})$ is identical to (1) but with the associated spectral polynomials $\left\{p_n^\pm(z)\right\}$ along with their recursion coefficients $\left\{\alpha_n^\pm,\beta_n^\pm\right\}$, and with the annihilation operators $a_\pm(E)$ and $a_\pm^j$ such that $\left[a_r(E),a_{r'}^\dagger(E')\right] = \delta_{r,r'}\,\delta(E-E')$ and $\left[a_r^i,(a_{r'}^j)^\dagger\right] = \lambda^{-1}\delta_{r,r'}\,\delta^{i,j}$ where $r$ and $r'$ stand for ±. The corresponding charged scalar antiparticle is represented by the following negative-energy quantum field

$$\bar{\Phi}(t,\vec{r}) = \frac{1}{\sqrt{2}}\left[\bar{\Psi}_+(t,\vec{r}) + \bar{\Psi}_-^\dagger(t,\vec{r})\right]. \tag{16b}$$

For this charged scalar particle, the Feynman propagator $\Delta_F(t'-t,\vec{r}\,'-\vec{r})$ between the two space-time points $(t,\vec{r})$ and $(t',\vec{r}\,')$ is constructed by combining the following two processes [1-4]:

(1) The creation of a particle from the vacuum $|0\rangle$ at $(t,\vec{r})$ and annihilating it later ($t'>t$) back into the vacuum at $(t',\vec{r}\,')$.

(2) The conjugate process of creating an antiparticle from the vacuum at $(t',\vec{r}\,')$ then annihilating it later ($t>t'$) at $(t,\vec{r})$.

That is,

$$\begin{gathered}\Delta_F(t'-t,\vec{r}\,'-\vec{r})=\langle 0|T\left(\bar{\Phi}(t',\vec{r}\,'),\Phi(t,\vec{r})\right)|0\rangle= \\ \langle 0|\bar{\Phi}(t',\vec{r}\,')\Phi(t,\vec{r})|0\rangle\theta(t'-t)+\langle 0|\Phi(t,\vec{r})\bar{\Phi}(t',\vec{r}\,')|0\rangle\theta(t-t')\end{gathered} \tag{17}$$

where $T$ is the time ordering operator and $\theta(x)=\left\{\begin{smallmatrix}1\ ,x>0\\ 0\ ,x<0\end{smallmatrix}\right.$. With the free propagator being determined, one needs to identify the type of interaction to account for the behavior of the scalar particle when coupled to its environment. Without such an interaction, the internal structure of the non-elementary particle (summation part of the quantum field) has no bearing on its free motion. Only in the presence of interaction will we observe the added effect of the internal structure. Moreover, the type and extent of such an effect will certainly depend on the nature of the interaction (electromagnetic, nuclear, gravitational, etc.). One way to incorporate the interaction of the particle with an external field is by using the gauge invariant minimal coupling scheme where the 4-gradient $\left(\partial_0,\vec{\nabla}\right)$ in the wave equation is replaced by $\left(\partial_0+igA_0,\vec{\nabla}+ig\vec{A}\right)$ with $g$ being the coupling parameter and $(A_0,\vec{A})$ the external 4-vector field. In Section 4, we present an example of nonlinear interaction in SAQFT and show how to calculate the scattering amplitudes to a given order in the coupling using Feynman diagrams. As in conventional QFT, the computed result of a physical process (like scattering) does not depend on the details of the kinematical construction of the free quantum fields but only on the main dynamical ingredients of the theory and on the interaction model. In SAQFT, this means that the results of a physical process depend only on the properties of the spectral polynomials $\{p_n(z)\}$ and on the type of interaction. Consequently, one should not worry much about the detailed construction of the spatial functions $\{\phi_n(\vec{r})\}$ beyond their completeness and that they belong to the solution space of the free wave equation which, for scalar particles, are guaranteed by Eq. (8) and Eq. (4), respectively.

In the Dirac-Coulomb problem, it was shown elsewhere [19,20] that the electromagnetic interaction of electrons, of which the Hydrogen atom is an example, is associated with the two-parameter Meixner-Pollaczek polynomial $P_n^\mu(z,\theta)$ where $\mu>0$ and $0<\theta<\pi$. This polynomial is known to have only a continuous spectrum. That is, the summation part in the orthogonality (6) is absent. Consequently, the internal structure is null and, in this case, SAQFT is equivalent to conventional QFT; both leading to QED. In Section 3, we present a simple nontrivial example where the particle structure will have an effect on the outcome.

For consistency, it is fruitful and possibly necessary to do dimensional analysis of the main objects in SAQFT in $n+1$ space-time. In the relativistic units $\hbar=c=1$, physical quantities like space, time, mass, field operators, etc. are measured in units of the energy; say $\mathcal{E}$ (e.g., MeV or GeV). For example, dimensional analysis shows that the annihilation/creation operators $a(\vec{k})$ and $a^{\dagger}(\vec{k})$ in conventional QFT are measured in units of $\mathcal{E}^{-n/2}$ whereas the field operators $\Psi(t,\vec{r})$ are measured in units of $\mathcal{E}^{(n-1)/2}$. On the other hand, the commutation relation (2) shows that the annihilation/creation operators $a(E)$, $a^{\dagger}(E)$, $a_j$ and $a_j^{\dagger}$ are measured in units of $\mathcal{E}^{-\frac{1}{2}}$. Moreover, equations (1) through (11) result in the following energy units for measuring the following objects in scalar SAQFT: $\phi_n(\vec{r}) \sim \mathcal{E}^{(n-1)/2}$, $f_0(E) \sim \mathcal{E}^{-\frac{1}{2}}$, $g_0(E_j) \sim \mathcal{E}^{\frac{1}{2}}$, $p_n(z) \sim \mathcal{E}^0$, $\psi(E,\vec{r}) \sim$ $\mathcal{E}^{(n/2)-1}$, and $\psi_j(\vec{r}) \sim \mathcal{E}^{n/2}$. Therefore, the field operator $\Psi(t,\vec{r})$ in scalar SAQFT is measured in units of $\mathcal{E}^{(n-1)/2}$ same as $\phi_n(\vec{r})$. However, for spinors (see Appendix B) the scale parameter $\lambda$ does not appear neither in the orthogonality (8a) nor in the completeness (8b) as seen in the corresponding equations (B9a) and (B9b). As a result, the dimensions of the spinor objects in SAQFT read as follows: $\phi_n(\vec{r}) \sim \mathcal{E}^{n/2}$, $\psi(E,\vec{r}) \sim \mathcal{E}^{(n-1)/2}$, and $\psi_j(\vec{r}) \sim \mathcal{E}^{(n+1)/2}$. Hence, the spinor field operator $\Psi(t,\vec{r})$ is measured in units of $\mathcal{E}^{n/2}$ same as $\phi_n(\vec{r})$.

It is worthwhile giving a snapshot in configuration space for the quantum field (1) in SAQFT at a given energy and comparing that to conventional QFT. For simplicity, we consider 1+1 space-time where such representation in QFT is proportional to $e^{ikx}$. For $E^2 > M^2$, this is just a sinusoidal oscillation of a fixed and finite amplitude that extends all the way to infinity. However, for $E^2 < M^2$, it is a decaying exponential with maximum at the origin and vanishing at the boundaries of space. On the other hand, in SAQFT such representation depends on the choice of spectral polynomials [defined by their three-term recursion relation (5) and orthogonality (6)] and the space functions [defined by their differential relation (4) and completeness (8)]. However, without any loss of generality, we find that a typical representation for $E^2 > M^2$ is oscillatory that extends all the way to infinity with an amplitude that varies locally but remains finite. For $E^2 < M^2$, the representation is also oscillatory but with an amplitude that decays rapidly and with a number of nodes equals to $j \in \{0,1,..,N\}$. Table 1 gives a pictorial comparison of a typical field in QFT and SAQFT. Nonetheless, in the absence of structure, QFT and SAQFT become equivalent as demonstrated by Eq. (7) above.

**Table 1**: Snapshot of a typical quantum field in 1+1 space-time at a given energy.

| Energy | QFT | SAQFT ( $j=8$ ) |
|---|---|---|
| $E^2 > M^2$ | 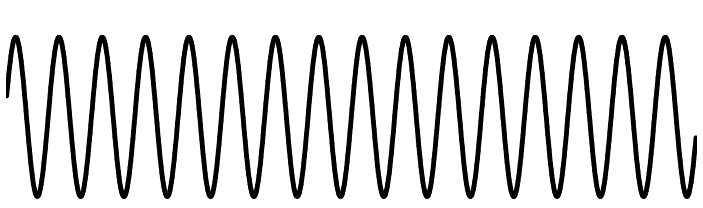 | 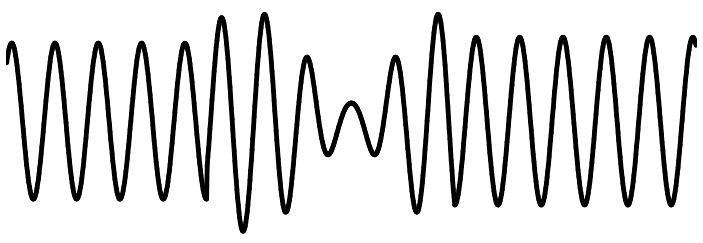 |
| $E^2 < M^2$ | 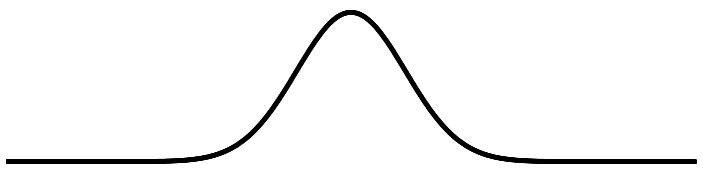 | 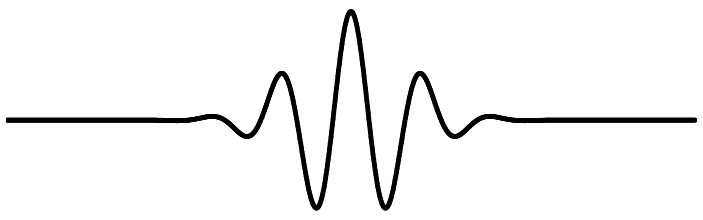 |

## 3. Scalar SAQFT example

As a simple but nontrivial example of scalar SAQFT, we consider a non-elementary scalar particle in 1+1 Minkowski space-time whose structure is associated with the three-parameter continuous dual Hahn polynomial $S_n^\mu(z;\sigma,\tau)$. The properties of this orthogonal polynomial that are relevant to our study are given in Appendix A. We consider here the special case where $\mu<0$ and $\sigma=\tau>-\mu$. From the formulation given above, it is obvious that the particle structure is determined by the recursion relation (5) *vis-à-vis* its recursion coefficients $\{\alpha_n,\beta_n\}$. The symmetric three-term recursion relation for the orthonormal version of the polynomial $S_n^\mu(z;\sigma,\sigma)$ is shown as Eq. (A6) in Appendix A giving

$$\alpha_n=\left(n+\mu+\sigma\right)^2+n\left(n+2\sigma-1\right)-\mu^2, \tag{18a}$$

$$\beta_n=-\left(n+\mu+\sigma\right)\sqrt{\left(n+1\right)\left(n+2\sigma\right)}. \tag{18b}$$

The physical effect of this internal structure becomes more evident if we write the differential equation satisfied by $\phi_n(x)$. We choose

$$\phi_n(x)=\sqrt{\Gamma(n+1)/\Gamma(n+2\sigma)}\,y^\sigma e^{-y/2}L_n^{2\sigma-1}(y), \tag{19}$$

where $y=e^{-\lambda x}$ and $L_n^{2\sigma-1}(y)$ is the Laguerre polynomial. The scale parameter $\lambda$ is real and positive with inverse length dimension and such that $\lambda\le|M/\mu|$. It could be considered as a measure of the size of the structure. The orthogonality of the Laguerre polynomials shows that $\{\phi_n(x)\}$ is an orthonormal set [i.e., $\bar{\phi}_n(x)=\phi_n(x)$]. Using the differential equation and recursion relation of the Laguerre polynomials, we obtain the free wave equation associated with this non-elementary scalar particle that replaces equation (4). It reads

$$\left[\frac{-1}{\lambda^2}\frac{d^2}{dx^2}+W(x)\right]\phi_n(x)=\alpha_n\phi_n(x)+\beta_{n-1}\phi_{n-1}(x)+\beta_n\phi_{n+1}(x), \tag{20}$$

where $W(x)$ is a manifestation of the structure of the particle corresponding to $S_n^\mu(z;\sigma,\sigma)$ and it reads

$$W(x)=\frac{1}{4}e^{-2\lambda x}+\left(\mu-\tfrac{1}{2}\right)e^{-\lambda x}. \tag{21}$$

Moreover, $z=(E^2-M^2)/\lambda^2$ and the size of the structure is equal to $N+1$, where $N$ is the largest integer less than $-\mu$. The rest of the objects needed to determine the quantum fields and propagators are the continuous and discrete components of the weight functions $\rho(z)$ and $\xi(z_j)$ in addition to the spectrum $\{z_j\}$ of the discrete structure. These are given in Appendix A by Eq. (A2), Eq. (A4) and Eq. (A5), respectively.

An interaction in this system is a physical process whereby the set of parameters $\{E,\lambda,\mu,\sigma\}$ that defines the particle is changed to $\{E',\lambda',\mu',\sigma'\}$. That is, $S_n^\mu(z;\sigma,\sigma)\mapsto$ $S_n^{\mu'}(z';\sigma',\sigma')$ which, for inelastic scattering, could alter the size and nature of the particle's structure. For example, if $\mu=-3.7\mapsto\mu'=-3.2$ while $\lambda$ remains the same then the structure will maintain its size $(N+1=4)$ but the levels of the structure will change from $E_j^2=M^2-\lambda^2(j-3.7)^2$ to $E_j'^2=M^2-\lambda^2(j-3.2)^2$ where $j=0,1,2,3$. However, if $\mu$ changes to, say $\mu'=-1.5$, then not only the levels will change but also the size of the structure (i.e., nature of the particle itself) changes from 4 to 2. Furthermore, if $\mu'$ becomes positive then the entire structure disappears signifying a total decay of the particle's constituents or structure due to the process.

# 4. Scattering in SAQFT

In this section, we give an illustrative example to outline the procedure for doing scattering calculation in SAQFT. For that, we consider the nonlinear interaction Lagrangian model $\mathscr{L}_I=\eta\otimes\Phi(\bar{\mathcal{X}}\mathcal{X})$, where $\Phi$ is a scalar, $\mathcal{X}$ is a spinor (see Appendix B), and $\eta$ is a dimensionless coupling tensor of rank three. The positive energy component of $\mathscr{L}_I$ reads as follows (for simplicity of the presentation, we show $\mathscr{L}_I$ for structureless neutral particles and for a single spinor component)

$$\mathscr{L}_I=\sum_{n,m,k=0}^{\infty}\eta_n^{m,k}\left[\int_{\Omega}e^{-\mathrm{i}(E-E'+E'')t}dzdz'dz''\sqrt{\rho(z)\omega(z')\omega(z'')}\right.$$
$$\left.p_n(z)q_m(z')q_k(z'')\phi_n(\vec{r})\bar{\vartheta}_m(\vec{r})\vartheta_k(\vec{r})a(E)b^\dagger(E')b(E'')\right] \tag{22}$$

The spectral polynomials $\{p_n(z)\}$ are associated with the scalar $\Phi$ whereas $\{q_n(z)\}$ are the spectral polynomials associated with the chosen spinor component of $\mathcal{X}$. Moreover, $\omega(z)$ is the continuous weight function associated with $\{q_n(z)\}$ and $\vartheta_n=\begin{pmatrix}\vartheta_n^+\\ \vartheta_n^-\end{pmatrix}$ is the corresponding four-component spinor basis functions that satisfy equations (B5) and (B9) in Appendix B. Just as in conventional QFT, it will become evident that the results of a physical process like the scattering amplitudes are independent of the details of the kinematical construction in the free theory. They depend only on the main dynamical ingredients of the theory, which in SAQFT are the spectral polynomials, and on the specific interaction model.

In conventional QFT, a physical process could be accounted for by summing all Feynman diagrams occurring within the process up to a given order. In SAQFT, however, the set of spectral polynomials associated with a given quantum field determines all physical properties of the corresponding particle. Thus, a scattering process in SAQFT is equivalent to the evaluation of the change in these polynomials. Now, the wave equation for the scalar field $\Phi(t,\vec{r})$ in the presence of interaction contains the nonlinear term $\eta\otimes(\bar{\mathcal{X}}\mathcal{X})$ added to the free Klein-Gordon equation resulting in a three-term recursion relation for $p_n(z)$ which differs from that of the free field relation (5) and reads

$$z\,p_n(z)=\alpha_n p_n(z)+\beta_{n-1}p_{n-1}(z)+\beta_n p_{n+1}(z)+\Delta p_n(z), \tag{23}$$

which is to be solved for the modified spectral polynomial. Henceforth, our task in this section is to calculate the added component $\Delta p_n(z)$ for the proposed model. For that, we use the Feynman diagrams to compute these changes up to second order in the coupling. In other scenarios, such change could also be evaluated perturbatively in powers of the physical parameters that appear in the recursion coefficients $\{\alpha_n\beta_n,\}$ or in the spectral parameter *z*. For example, in Section 3, we could have $|\lambda/M| << 1$ or $\sigma+\mu << 1$, and so on. Ideally, if we could evaluate $\Delta p_n(z)$ to all orders then we should obtain $\Delta p_n(z)=\Delta\alpha_n(\eta)p_n(z)+\Delta\beta_{n-1}(\eta)p_{n-1}(z)+\Delta\beta_n(\eta)p_{n+1}(z)$ and end up with one of two cases. If the scattering is elastic then the particle will maintain its identity, which means that the modified spectral polynomial will satisfy the same recursion relation but with possibly different initial value $p_1(z)$ (i.e., different spin, polarization, etc.). This means that for elastic scattering, $\Delta p_n(z)$ is a two-parameter linear function of *z* multiplied by $\delta_{n,0}$. However, if the scattering is inelastic then the particle changes its identity, which means that the modified spectral polynomial will satisfy a different recursion relation with the modified coefficients $\{\alpha'_n,\beta'_n\}=\{\alpha_n+\Delta\alpha_n,\beta_n+\Delta\beta_n\}$. Below, we outlined a perturbative procedure to calculate $\Delta p_n(z)$ for the proposed nonlinear interacting model up to second order by using Feynman diagrams. However, the rules of the diagrams used in conventional QFT must be amended to be suitable for application in SAQFT.

Figure 1 gives a graphical representation for the evaluation of $\Delta p_n(z)$ to second order by following the rules of Feynman diagrams for this specific model. However, these rules which are well established in conventional QFT, need few alterations to be suitable for use in SAQFT. For example, in conventional QFT, propagators in the diagrams are associated with the linear momentum vectors $\vec{k}$ making vector addition very critical in closed loops. In SAQFT, however, the spectral parameters (arguments of the spectral polynomials) and polynomial degrees are assigned to propagators instead, making spectral parameter *addition* and *subtraction* in closed loops critical. Energy-momentum conservation at the vertices makes adding and subtracting these spectral parameters a nontrivial operation. Shortly, we will show how this is done. Other alterations of the rules might be necessary as the theory becomes more developed.

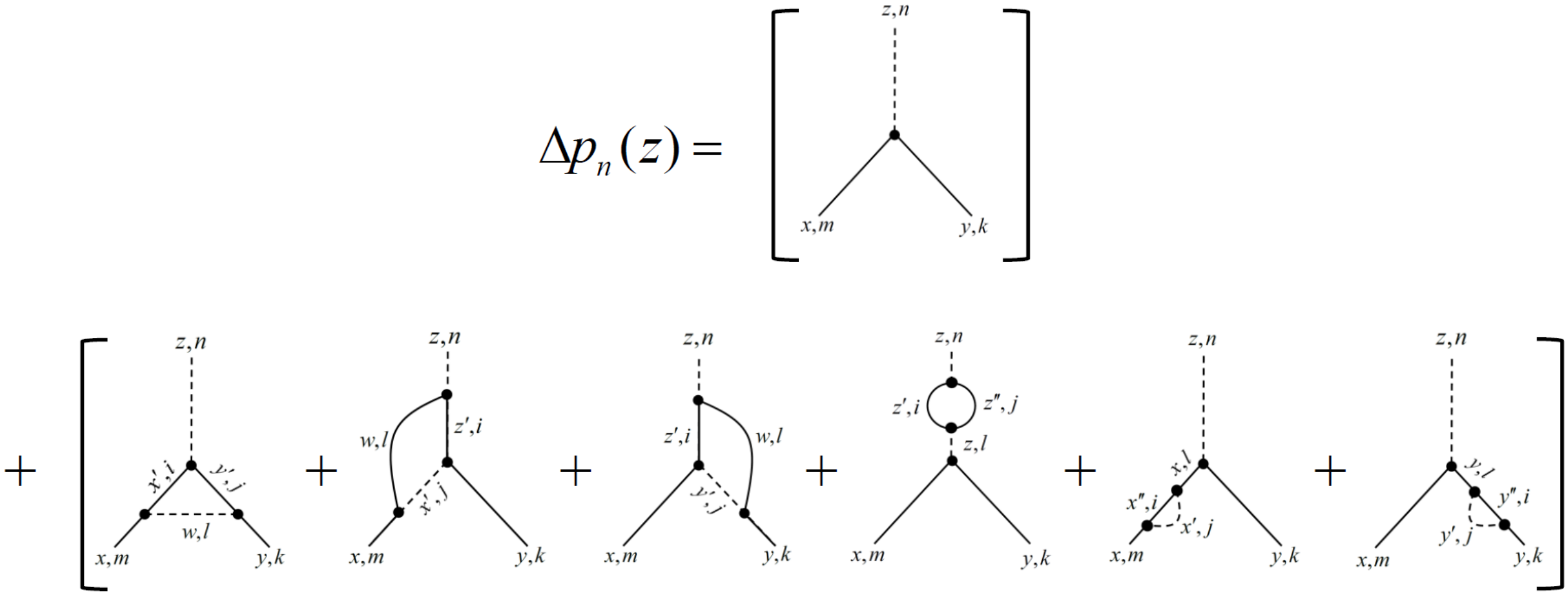

**Fig. 1**: Feynman diagrams contributing to the evaluation of $\Delta p_n(z)$ in the model up to 2nd order in the coupling. The solid/dashed traces correspond to the spinor/scalar propagators. The inflow

of the two spinor particle beams $(x,m)$ and $(y,k)$ is from the bottom whereas the outflow of the scalar particle beam $(z,n)$ is upwards. All loops are taken in the counterclockwise direction.

In Figure 1, the pair $(a,l)$ on a Feynman propagator (or at one end for incoming/outgoing particles) stands for the associated spectral polynomial $p_l(a)$ or $q_l(a)$, with $a$ being the spectral parameter (continuous and/or discrete). The two incoming spinor particles are associated with the spectral polynomials $q_m(x)$ and $q_k(y)$. The outgoing scalar particle is associated with the spectral polynomial $p_n(z)$. Therefore, in this scattering process, the degrees $\{m,k,n\}$ and associated spectral parameters $\{x,y,z\}$ are fixed. At each vertex, the energy-momentum vector is conserved. For example, in the first diagram on the second line of the equation and with the choice of a counterclockwise loop, the three energy conservation equations are: $E(x)=E(w)-E(x')$, $E(y)=E(y')-E(w)$, and $E(z)=E(y')-E(x')=E(x)+E(y)$. This leads to a special rule for adding and subtracting spectral parameters in the Feynman diagrams that goes as follows. We define $x\oplus y$ as the quantity $z$ that makes $E(x)+E(y)=E(z)$. Similarly, $x\ominus y:=z$ means that $E(x)-E(y)=E(z)$. Alternatively, we can write $x\oplus y=E^{-1}\left(E(x)+E(y)\right)$ and $x\ominus y=E^{-1}\left(E(x)-E(y)\right)$. With $[E(z)]^2=z+M^2$, this operation is nontrivial but becomes simpler for massless particles where $x\oplus y=x+y+2\,\mathrm{sgn}\sqrt{xy}$ and $x\ominus y=x+y-2\,\mathrm{sgn}\sqrt{xy}$ with "sgn" being the sign of $E(x)E(y)$. If both $x$ and $y$ are positive (i.e., in the continuous spectrum Ω) then $x\oplus y$ is either:

(i) Positive (belonging to Ω) and greater than or equal to $3M^2$ if the particles associated with $x$ and $y$ are both with positive or both with negative energies, or

(ii) Having indefinite sign (but, of course, always greater than or equal to $-M^2$) if the particles associated with $x$ and $y$ have energies of different signs.

The reverse is true for $x\ominus y$. Table 2 shows all possible range of values of $x\oplus y$ and $x\ominus y$. A remarkable byproduct of working with the spectral parameters in SAQFT is a novel algebraic system defined and detailed in Ref. [21].

**Table 2**: Range of values of $z=x\oplus y$. For $z=x\ominus y$, the ± in the first column must be interchanged.

| sgn | $(x,y)\geq 0$ | $x\geq 0, -M^2\leq y\leq 0$ | $y\geq 0, -M^2\leq x\leq 0$ | $-M^2\leq (x,y)\leq 0$ |
|---|---|---|---|---|
| + | $z\geq 3M^2$ | $z\geq 0$ | $z\geq 0$ | $-M^2\leq z\leq 3M^2$ |
| − | $z\geq -M^2$ | $z\geq -M^2$ | $z\geq -M^2$ | $-M^2\leq z\leq 0$ |

If we combine the continuous and discrete weight functions in the generalized orthogonality (6) as $\tilde{\rho}(z):=\rho(z)+\xi(z)\sum_{j=0}^{N}\delta(z-z_j)$, then the orthogonality takes the following compact form

$$\int_{\tilde{\Omega}} \tilde{\rho}(z) p_n(z) p_m(z) dz = \delta_{n,m}, \tag{24}$$

where $\tilde{\Omega} = \Omega \cup [E_0, E_N]$. Now, the quantum field representation (1) means that each propagator in the Feynman diagrams with the pair $(a,l)$ is associated not just with $p_l(a)$ but, in fact, with $f_0\left(E(a)\right) p_l(a)$ for the continuous spectrum and with $g_0\left(E(a_j)\right) p_l(a_j)$ for the discrete, which we can write collectively as $\sqrt{\tilde{\rho}(a)}\, p_l(a)$ for the scalar and $\sqrt{\tilde{\omega}(a)}\, q_l(a)$ for the spinor. The first line of the equation in Figure 1 gives $\Delta p_n(z)$ to first order as $\Delta p_n(z) = \eta_n^{m,k}$, which is the bare vertex. On the other hand, the first diagram on the second line of the equation with the choice of counterclockwise loop gives the following second order contribution to $\Delta p_n(z)$

$$\begin{aligned} &\sqrt{\tilde{\omega}(w \ominus x)\tilde{\omega}(y \oplus w)\tilde{\omega}(w \ominus x)\tilde{\rho}(w)\tilde{\omega}(y \oplus w)\tilde{\rho}(w)} \\ &\times \sum_{i,j,l=0}^{\infty} \eta_n^{i,j} \eta_m^{i,l} \eta_k^{j,l} q_i(w \ominus x) q_j(y \oplus w) q_i(w \ominus x) p_l(w) q_j(y \oplus w) p_l(w) \end{aligned} \tag{25}$$

Using the property $y \oplus w = w \oplus y$ and collecting terms, we can rewrite (25) as follows

$$\tilde{\omega}(w \ominus x)\tilde{\omega}(w \oplus y)\tilde{\rho}(w) \sum_{i,j,l=0}^{\infty} \eta_n^{i,j} \eta_m^{i,l} \eta_k^{j,l} q_i^2(w \ominus x) q_j^2(w \oplus y) p_l^2(w), \tag{26}$$

The second diagram on the second line of the equation in the figure with a counterclockwise loop gives the following second order contribution to $\Delta p_n(z)$

$$\tilde{\omega}(w \oplus z)\tilde{\rho}(w \oplus x)\tilde{\omega}(w) \sum_{i,j,l=0}^{\infty} \eta_n^{i,l} \eta_m^{j,l} \eta_k^{i,j} q_i^2(w \oplus z) p_j^2(w \oplus x) q_l^2(w), \tag{27}$$

The third diagram on the second line of the equation is topologically equivalent to the second giving a contribution identical to (27) but with $x \leftrightarrow y$ and $m \leftrightarrow k$. The fourth diagram, on the other hand, gives the following contribution to $\Delta p_n(z)$ with a counterclockwise loop

$$\tilde{\omega}(w)\tilde{\omega}(w \oplus z) \sum_{i,j,l=0}^{\infty} \eta_n^{i,j} \eta_l^{i,j} \eta_l^{m,k}\, q_i^2(w) q_j^2(w \oplus z). \tag{28}$$

where we have set $w := z'$. The fifth diagram on the second line of the equation in the figure gives the following second order contribution to $\Delta p_n(z)$

$$\tilde{\rho}(w)\tilde{\omega}(w \oplus x) \sum_{i,j,l=0}^{\infty} \eta_m^{i,j} \eta_l^{i,j} \eta_l^{n,k}\, p_j^2(w) q_i^2(w \oplus x), \tag{29}$$

where we have set $w := x''$. The last diagram on the second line of the equation is topologically equivalent to the fifth and gives a contribution identical to (29) but with $x \leftrightarrow y$ and $m \leftrightarrow k$.

As the two incoming particles being already selected and prepared in the lab (e.g., in a monochrome scattering experiment) then their corresponding energies $\{E(x), E(y)\}$, spectral parameters $\{x, y\}$ and spectral polynomial degrees $\{m, k\}$ are fixed. That is, $q_m(x)$ and $q_k(y)$ are fixed. Likewise, the output channel for this scattering process has also been determined to correspond to $p_n(z)$, where $E(z) = E(x) + E(y)$. Hence, the spectral parameter $z$ and associated polynomial degree $n$ are also fixed. Therefore, $w$ and the degrees $\{i, j, l\}$ become the only variables in Figure 1. However, since the infinite summations in (26) to (29) have already accounted for all possible degrees $\{i, j, l\}$, we only need to integrate over the whole possible range of values of $w$ (continuous and discrete). That is, integrating $w$ over the complete energy domain $\tilde{\Omega}$ (integration over $\Omega$ and summation over $\{w_i\}_{i=0}^N$). Therefore, at this stage in the process we need to perform this integration-summation task in order to write down the one-loop vertex correction shown in Figure 1. We start by defining the following "fundamental SAQFT integrals":

$$\zeta_n^p := \int_{\tilde{\Omega}} \tilde{\rho}(w) p_n^2(w)\, dw = \int_{\Omega} \rho(w) p_n^2(w)\, dw + \sum_{j=0}^{N} \xi(w_j) p_n^2(w_j) = 1, \tag{30a}$$

$$\begin{aligned}\zeta_{n,m}^{p,p}(a) &:= \int_{\tilde{\Omega}} \tilde{\rho}(w)\tilde{\rho}(w \circledast a) p_n^2(w) p_m^2(w \circledast a)\, dw \\ &= \int_{\Omega} \rho(w)\rho(w \circledast a) p_n^2(w) p_m^2(w \circledast a)\, dw + \sum_{j=0}^{N} \xi(w_j)\xi(w_j \circledast a) p_n^2(w_j) p_m^2(w_j \circledast a)\end{aligned} \tag{30b}$$

$$\begin{aligned}\zeta_{n,m}^{p,q}(a) &:= \int_{\tilde{\Omega}} \tilde{\rho}(w)\tilde{\omega}(w \circledast a) p_n^2(w) q_m^2(w \circledast a)\, dw \\ &= \int_{\Omega} \rho(w)\omega(w \circledast a) p_n^2(w) q_m^2(w \circledast a)\, dw + \sum_{j=0}^{N} \xi(w_j)\varsigma(w_j \circledast a) p_n^2(w_j) q_m^2(w_j \circledast a)\end{aligned} \tag{30c}$$

$$\begin{aligned}\zeta_{n,m}^{q,q}(a) &:= \int_{\tilde{\Omega}} \tilde{\omega}(w)\tilde{\omega}(w \circledast a) q_n^2(w) q_m^2(w \circledast a)\, dw \\ &= \int_{\Omega} \omega(w)\omega(w \circledast a) q_n^2(w) q_m^2(w \circledast a)\, dw + \sum_{j=0}^{N} \varsigma(w_j)\varsigma(w_j \circledast a) q_n^2(w_j) q_m^2(w_j \circledast a)\end{aligned} \tag{30d}$$

$$\begin{aligned}\zeta_{n,m,k}^{p,p,p}(a,b) &:= \int_{\Omega} \rho(w)\rho(w \circledast a)\rho(w \circledast b) p_n^2(w) p_m^2(w \circledast a) p_k^2(w \circledast b)\, dw \\ &+ \sum_{j=0}^{N} \xi(w_j)\xi(w_j \circledast a)\xi(w_j \circledast b) p_n^2(w_j) p_m^2(w_j \circledast a) p_k^2(w_j \circledast b)\end{aligned} \tag{30e}$$

$$\begin{aligned}\zeta_{n,m,k}^{p,p,q}(a,b) &:= \int_{\Omega} \rho(w)\rho(w \circledast a)\omega(w \circledast b) p_n^2(w) p_m^2(w \circledast a) q_k^2(w \circledast b)\, dw \\ &+ \sum_{j=0}^{N} \xi(w_j)\xi(w_j \circledast a)\varsigma(w_j \circledast b) p_n^2(w_j) p_m^2(w_j \circledast a) q_k^2(w_j \circledast b)\end{aligned} \tag{30f}$$

$$\begin{aligned}\zeta_{n,m,k}^{p,q,q}(a,b) &:= \int_{\Omega} \rho(w)\omega(w \circledast a)\omega(w \circledast b) p_n^2(w) q_m^2(w \circledast a) q_k^2(w \circledast b)\, dw \\ &+ \sum_{j=0}^{N} \xi(w_j)\varsigma(w_j \circledast a)\varsigma(w_j \circledast b) p_n^2(w_j) q_m^2(w_j \circledast a) q_k^2(w_j \circledast b)\end{aligned} \tag{30g}$$

$$\zeta^{q,q,q}_{n,m,k}(a,b) := \int_{\Omega} \omega(w)\omega(w \circledast a)\omega(w \circledast b) q_n^2(w) q_m^2(w \circledast a) q_k^2(w \circledast b)\, dw$$
$$+ \sum_{j=0}^{N} \varsigma(w_j)\varsigma(w_j \circledast a)\varsigma(w_j \circledast b) q_n^2(w_j) q_m^2(w_j \circledast a) q_k^2(w_j \circledast b) \qquad \text{(30h)}$$

where $\circledast$ stands for either $\oplus$ or $\ominus$ and $\varsigma(z_j)$ is the discrete weight function associated with the spectral polynomials $\{q_n(z)\}$. These are the lowest three in the series of fundamental SAQFT integrals corresponding a single closed loop with one, two, and three vertices, respectively. Note that the integrals (30b), (30e), (30f), and (30h) do not contribute in the present model. The orthogonality (24) shows that the first integral is, in fact, trivial. It corresponds to the bare vertex in the first diagram of Figure 1. The integrals in (30b)-(30h) are valid only if $w \circledast a$ and $w \circledast b$ are elements of the continuous spectrum Ω. Table 2 is useful in making such validity determination. For example, if $\{x,y,z,w\}$ belong to the continuous spectrum Ω and thus positive, the quantity $w \ominus x$ in the integral with $E(x) \gtrless 0$ becomes positive definite (in fact, greater than or equal to $3M^2$) by choosing the (arbitrary) positive variable *w* such that $E(w) \lessgtr 0$, respectively. The summations, on the other hand, are valid only if $w_i \circledast a_j$ and $w_i \circledast b_j$ are elements of the discrete spectrum set $\{w_i\}_{i=0}^{N}$. For higher order loops with multiple integrals over several arbitrary spectral variables (e.g., $w, u, v, ..,$ etc.), it is always possible to choose, say *u* and *w*, such that $w \circledast u \in \Omega$, $w_i \circledast u_j \in \{w_i\}_{i=0}^{N}$ and $u_i \circledast w_j \in \{u_i\}_{i=0}^{N}$. Now, the orthogonality relation (6), or equivalently (24), guarantees that the integrals in (30b)-(30h) are finite provided that the range of values of *w* is such that $w \circledast a$ and $w \circledast b$ belong to Ω [22]. Moreover, one can show that these integrals are not just finite but, in fact, fall within the interval $[0,+1]$. Additionally, in the limit as *n, m,* and/or *k* go to infinity, these integrals go to zero. These properties will be demonstrated below in a numerical example. Finally, after carrying out the integration and summation in (26) through (29), we obtain the following second order correction to the vertex diagram shown in Figure 1

$$\Delta p_n(z) \approx \eta_n^{m,k} + \left[\vartheta^{p,q,q}_{n,m,k}(x,y) + \vartheta^{q,q,p}_{k,m,n}(x,z) + \vartheta^{q,q,p}_{m,k,n}(y,z)\right]$$
$$+ \left[\hat{\vartheta}^{q,q}_{n,m,k}(z) + \hat{\vartheta}^{p,q}_{m,n,k}(x) + \hat{\vartheta}^{q,p}_{k,n,m}(y)\right] \qquad \text{(31)}$$

where $\vartheta^{p,q,q}_{n,m,k}(a,b) := \sum_{i,j,l=0}^{\infty} \eta_n^{i,j}\eta_m^{i,l}\eta_k^{j,l}\zeta^{p,q,q}_{l,i,j}(a,b)$, $\hat{\vartheta}^{p,q}_{n,m,k}(a) := \sum_{i,j,l=0}^{\infty} \eta_n^{i,j}\eta_l^{i,j}\eta_l^{m,k}\zeta^{p,q}_{i,j}(a)$ and we assumed that $\eta_n^{m,k} = \eta_n^{k,m}$. These infinite sums converge because the vertex parameters $\{\eta_n^{m,k}\}$ are finite and $\zeta^{p,q}_{i,j}(a)$ as well as $\zeta^{p,q,q}_{l,i,j}(a,b)$ go to zero fast enough as their indices tend to infinity.

Evidently, the evaluation of the integrals of the type shown in (30) is of prime importance in the calculation of the Feynman diagrams in SAQFT. We find that Gauss quadrature integral approximation associated with the spectral polynomials $\{p_n(w)\}$ and/or $\{q_n(w)\}$ produces highly accurate results. For a brief description of Gauss quadrature, one may consult, for example, Ref. [23] or Ref. [24]. In summary, it goes as follows. Let *K* be a large-enough integer (called the order of the quadrature) and let *J* be the $K \times K$ truncated version of the tridiagonal symmetric matrix *Q* shown below in Eq. (42), which is the Jacobi matrix associated with the spectral polynomials $\{p_n(w)\}$. We designate the $K$ real eigenvalues of $J$ as the set $\{\lambda_k\}_{k=0}^{K-1}$ and their corresponding normalized eigenvectors as $\{\Lambda_{i,k}\}_{i=0}^{K-1}$. In this setting, Gauss quadrature gives the following approximation for the integrals in (30)

$$\zeta_{n,m}^{p,p}(a) \cong \sum_{k=0}^{K-1} \Lambda_{n,k}^2\, \rho(\lambda_k \circledast a)\, p_m^2(\lambda_k \circledast a)\,, \tag{32a}$$

$$\zeta_{n,m}^{p,q}(a) \cong \sum_{k=0}^{K-1} \Lambda_{n,k}^2\, \omega(\lambda_k \circledast a) q_m^2(\lambda_k \circledast a)\,, \tag{32b}$$

$$\zeta_{n,m,l}^{p,p,p}(a,b) \cong \sum_{k=0}^{K-1} \Lambda_{n,k}^2\, \rho(\lambda_k \circledast a)\rho(\lambda_k \circledast b)\, p_m^2(\lambda_k \circledast a)\, p_l^2(\lambda_k \circledast b)\,. \tag{32c}$$

$$\zeta_{n,m,l}^{p,p,q}(a,b) \cong \sum_{k=0}^{K-1} \Lambda_{n,k}^2\, \rho(\lambda_k \circledast a)\omega(\lambda_k \circledast b)\, p_m^2(\lambda_k \circledast a) q_l^2(\lambda_k \circledast b)\,. \tag{32d}$$

$$\zeta_{n,m,l}^{p,q,q}(a,b) \cong \sum_{k=0}^{K-1} \Lambda_{n,k}^2\, \omega(\lambda_k \circledast a)\omega(\lambda_k \circledast b) q_m^2(\lambda_k \circledast a) q_l^2(\lambda_k \circledast b)\,. \tag{32e}$$

The approximation improves with the order of the quadrature, $K$. Figure 2 is a sample plot of $\zeta_{n,m}^{p,p}(a)$, with $p_n(w)$ being the normalized version of the Laguerre polynomial $L_n^\nu(w)$ and $\circledast \mapsto \oplus$, $E(w)E(a) > 0$. The figure demonstrates the diminishing value of $\zeta_{n,m}^{p,p}(a)$ with increasing $n$ and $m$.

Obtaining finite values for these fundamental SAQFT integrals that represent closed loops in the Feynman diagrams is a remarkable property of the theory that will certainly have a positive impact on renormalization. At best, this property (if sustained for all loops) may eliminate the need for renormalization altogether rending the theory finite. However, proving such claim of novelty remains an open problem [22].

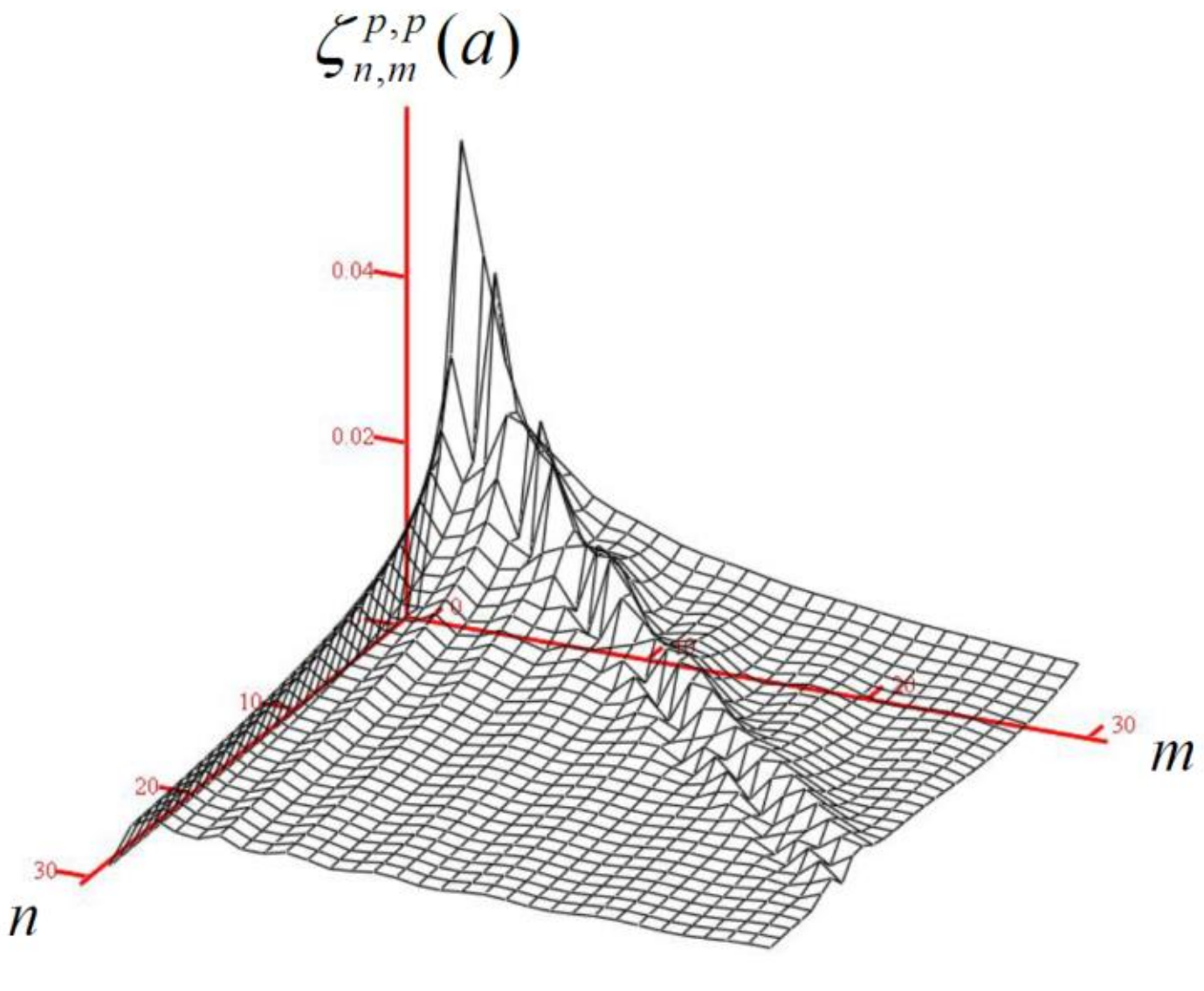

**Fig. 2**: Plot of $\zeta_{n,m}^{p,p}(a)$ with $p_n(w)$ being the normalized Laguerre polynomial $L_n^\nu(w)$, $\circledast \mapsto \oplus$, and $E(w)E(a) > 0$. We took $M = 1.0$, $a = 2.0$, $\nu = 1.5$ and the quadrature order $K = 200$.

The effect of the same interaction model, $\mathscr{L}_I = \eta \otimes \Phi(\bar{\mathcal{X}}\mathcal{X})$, on the scattering of the spinor particle can also be obtained perturbatively by following the same procedure. That is, we interpret Fig. 1 as incoming scalar particle beam corresponding to the spectral polynomials $p_n(z)$ and an incoming spinor particle beam corresponding to $q_m(x)$ and then we calculate the deviation $\Delta q_k(y)$ in the recursion relation associated with the spectral polynomial $q_k(y)$ for the outgoing spinor beam.

For completeness, we end this section by calculating the first order correction to the propagator (self-energy) in this model where the corresponding Feynman diagram for the spinor is shown as Figure 3. Writing $w := x''$, we obtain the following first order contribution to $\Delta q_n(x)$

$$\tilde{\omega}(w)\tilde{\rho}(w \ominus x)\sum_{i,j=0}^{\infty} \eta_n^{i,j}\eta_m^{i,j}\, q_i^2(w)\, p_j^2(w \ominus x)\,. \tag{33}$$

After integrating the arbitrary energy variable $w$ over $\tilde{\Omega}$, we obtain the following self-energy correction to first order as shown in Figure 3

$$\Delta q_n(x) \approx \sum_{i,j=0}^{\infty} \eta_n^{i,j}\eta_m^{i,j}\zeta_{i,j}^{q,p}(x)\,. \tag{34}$$

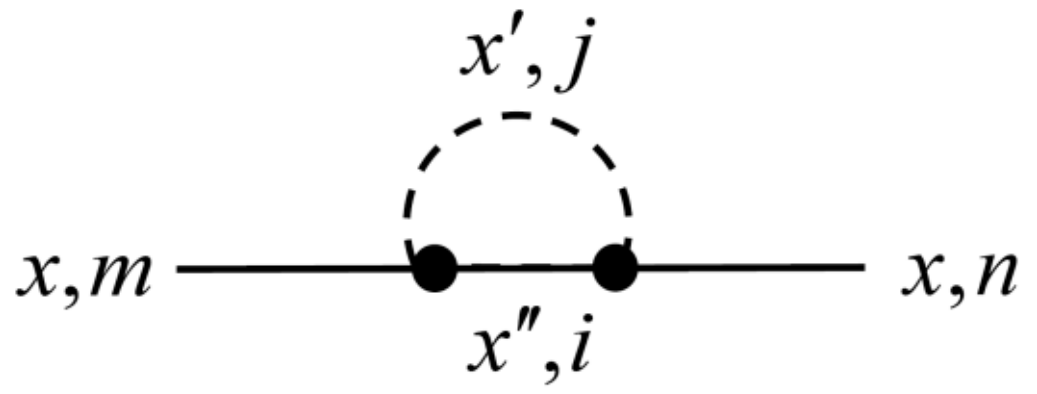

**Fig. 3**: Spinor self-energy correction diagram to first order in the coupling.

Similar treatment gives the following self-energy correction to the scalar propagator shown in Figure 4

$$\Delta p_n(x) \approx \sum_{i,j=0}^{\infty} \eta_n^{i,j}\eta_m^{i,j}\zeta_{i,j}^{q,q}(x)\,. \tag{35}$$

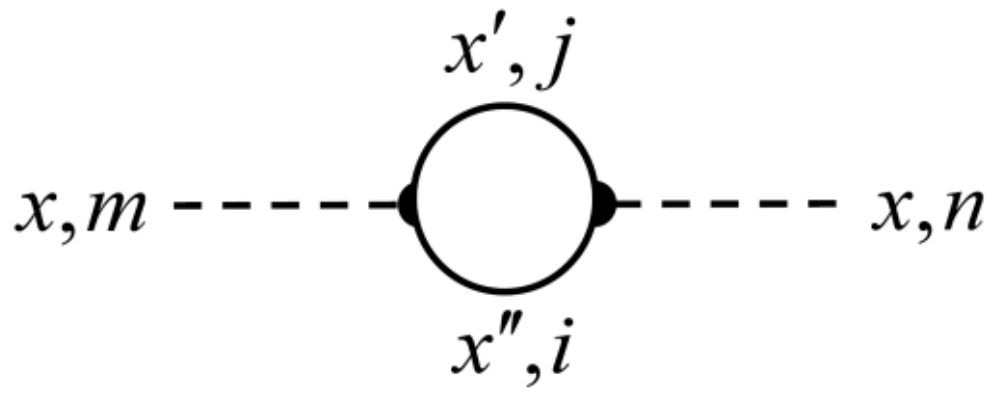

**Fig. 4**: Scalar self-energy correction diagram to first order in the coupling.

## 5. Conclusions and comments

In this work, we posed the critical question: Why doesn't conventional QFT work well at *very low energies* for particles with structure (e.g., in nuclear physics with hadrons)? Instead of laying blame on strong coupling and relying on the "asymptotic freedom" at higher energies, we suggested a simple answer: it is because of a missing piece; the particle structure itself. Therefore, we adopted the view that instead of assuming structureless particles at a given energy only to discover that at higher energy they do have structure, is to develop a practical QFT with structure already built-in. Consequently, we introduced an algebraic version of QFT that accommodates particles with structure, which is resolved in the energy domain. The theory of orthogonal polynomials plays a critical role in the formulation. It is hoped that the brief outline of the theory given here will motivate further studies using this approach towards a more effective and generalized QFT. Incorporating the particle structure may bring new elements into the theory that could be exploited to tackle some of the persistent difficulties in conventional QFT. We believe that these new elements may have a positive impact on the renormalization program (e.g., the size of the structure provides a natural cutoff of ultraviolet divergences). Moreover, energy integration (both continuous and discrete) over closed loops in the Feynman diagrams could render such integrals finite by exploiting the orthogonality relation (6) as demonstrated in the interaction model of Section 4.

In Appendices B and C, we give a brief mathematical presentation of SAQFT for spinor particles with structure and for the massless (structureless) vector field (e.g., the electromagnetic vector potential). Moreover, in Appendix D we provide the SAQFT formulation for the massive vector boson.

Recently, we proposed a mechanism within SAQFT whereby massless particles can acquire mass by a special nonlinear coupling to a universal massless scalar field [25]. This is analogous to the Higgs mechanism in the standard model of elementary particles in conventional QFT. However, in SAQFT, the universal massless scalar (the analogous to the Higgs particle) remains massless. If this proves successful, it could also provide a resolution to the problem of the missing dark energy since no massless scalar particle has ever been detected. Consequently, it would be interesting to devise a process within SAQFT that gives a unique signature for the universal scalar particle and design an experimental setup to detect such signature.

One of the remaining tasks in SAQFT (as presented here) is to establish the relativistic invariance of its physical objects and covariance of its elements under the space-time Poincaré transformation (Lorentz rotation + translation). It is conceivable that results from such an endeavor may alter and/or improve on the development of the theory as presented in this work. Another task is the development of the non-abelian version of SAQFT and the corresponding quantum Yang-Mills theory. Nonetheless, a practical initial study is to reproduce some of the well-known results in QED. The outline of such a study is given in Appendix C.

Finally, we conclude with the following ten-point remarks:

- The SAQFT presented in this work does not require specifying a free field wave equation. In fact, the formulation of the theory is built upon five defining postulates:

    (i) A physical system is represented by a quantum field and its conjugate that are resolved in the energy domain which, for scalar particles, are given by Eq. (1) and Eq. (9), respectively.

(ii) The creation/annihilation operators in the quantum field satisfy a commutation algebra which, for scalar particles, is given by relations (2).

(iii) All physical properties of the system are determined by the spectral polynomials satisfying the three-term recursion relation (5) and generalized orthogonality (6).

(iv) The completeness of spatial functions in the expansion series of the continuous and discrete Fourier kernels which, for scalar particles, is given by Eq. (8) and satisfying Eq. (4).

(v) Interaction in the theory results from a tensor coupling among the spectral polynomials of the interacting quantum fields.

The objects of the postulates in the case of spinors are shown in Appendix B by Eq. (B2) and Eq. (B10) for the quantum fields, by the anti-commutation algebra (B3) for the creation/annihilation operators, and by Eq. (B5) for the spatial spinor functions. In their classic book on QFT [1], Bjorken and Drell posed a fundamental question at the beginning of Section 11.1 on page 4: "*…why local field theories, that is, theories of fields which can be described by differential laws of wave propagation, have been so extensively used and accepted*." The answer given therein declares that "*there exist no convincing form of a theory which avoids differential field equations*." It is expected that the theory proposed here where the differential wave equation is replaced by the equivalent algebraic three-term recursion relation constitutes a viable alternative.

- One may suggest a modification in conventional QFT to incorporate particle structure in a simpler manner without going through the sumptuous construction of SAQFT involving orthogonal polynomials. That is, the conventional quantum field for scalar particles with structure could have been redefined as follows

$$\tilde{\Psi}(t,\vec{r}) = \int_{\vec{k}\in\mathbb{R}^3} e^{-\mathrm{i}Et+\mathrm{i}\vec{k}\cdot\vec{r}}\, a(\vec{k})\frac{d^3k}{\sqrt{(2\pi)^3 2E}} + \sum_{j=0}^{N} e^{-\mathrm{i}E_j t-|\vec{k}_j\cdot\vec{r}|} a_j\,. \tag{36}$$

However, there is a dimensional mismatch between the integral and sum unless we propose that $\left[a_i, a_j^\dagger\right] = \upsilon\delta_{i,j}$, where the parameter $\upsilon$ has the dimension of energy squared (i.e. $\upsilon \propto M^2$, $\upsilon \propto \lambda^2$ or $\upsilon \propto E_j^2$, etc.). Moreover, the representation (36) results in a quantum field that is "over-complete". Namely, $\left\|\tilde{\Psi}(t,\vec{r})\right\| > 1$ because the continuous spectrum is already complete since $\left\|e^{\mathrm{i}\vec{k}\cdot\vec{r}}\right\| = 1$. On the other hand, the orthogonality (6) of the spectral polynomials in SAQFT gives the completeness of the quantum field $\Psi(t,\vec{r})$ in (1) as a sum of the continuous and discrete spectra that could be written symbolically as follows

$$\left\|\Psi(t,\vec{r})\right\| = \left\|\psi(E,\vec{r})\right\| + \sum_{j=0}^{N}\left\|\psi_j(\vec{r})\right\| = 1\,. \tag{37}$$

In fact, even for very small structures, where for example $N = 0$ or 1, it may happen that the contribution of the structure [i.e., the sum in Eq. (37)] could dominate completeness. Another way to see the inadequacy of the representation (36) is to note that, in the language of SAQFT, the expansion of $e^{\mathrm{i}\vec{k}\cdot\vec{r}}$ is associated with the Hermite polynomial (or the Gegenbauer polynomial) whose entire spectrum is continuous leaving no room for including an internal discrete structure [*cf.* Eq. (7)].

- If the set of spectral polynomials $\{p_n(z)\}$ is endowed *only* with a discrete spectrum, then the continuous integral in the definition of the quantum field (1) does not appear and the orthogonality (6) consists only of the summation part. Examples of such polynomials include the Meixner, Charlier, dual Hahn, Krawtchouk and the Racah polynomials [26]. We conjecture that such systems might constitute a viable alternative to the traditional approach that accounts for the confinement of particles like the quarks. They could also be used in describing point contact (zero range) interactions (those where massless gauge bosons that mediate coupling in the interaction are absent). However, if such discrete polynomials are *finite*, i.e., $\{p_n(z_j)\}_{n,j=0}^N$, which may not be appropriate for representing quantum fields with infinite degrees of freedom, then the associated configuration space functions must form a complete finite discrete set, $\{\phi_n(\vec{r}_j)\}_{n,j=0}^N$, over the space lattice $\{\vec{r}_j\}_{j=0}^N$. That is,

$$\sum\nolimits_{n=0}^N \phi_n(\vec{r}_i)\bar{\phi}_n(\vec{r}_j) = \sum\nolimits_{n=0}^N \bar{\phi}_n(\vec{r}_i)\phi_n(\vec{r}_j) = l^{-2}\,\delta_{i,j}\,, \tag{38a}$$

$$\sum\nolimits_{j=0}^N \varsigma_j \phi_n(\vec{r}_j)\bar{\phi}_m(\vec{r}_j) = \sum\nolimits_{j=0}^N \varsigma_j \bar{\phi}_n(\vec{r}_j)\phi_m(\vec{r}_j) = l\,\delta_{n,m}\,, \tag{38b}$$

  for some positive discrete weight $\varsigma_j$ and lattice size $\lambda^{-1} = l$. For example, in 1+1 space-time with a linear lattice of total length $L$ and with $l = L/N$, we could take

$$\phi_n(x_j) = \bar{\phi}_n(x_j) = (N!)\sqrt{\frac{\vartheta^{n+j}(1-\vartheta)^{N-n-j}}{n!\,j!\,(N-n)!\,(N-j)!}}\;{}_2F_1\!\left(\left.{}^{-n,-j}_{\;-N}\right|\vartheta^{-1}\right), \tag{39}$$

  where $x_j = jl$, $0 < \vartheta < 1$, and $\varsigma_j = l$. For a proof of completeness and orthogonality of these discrete functions, see Appendix A in Ref. [27] for the Krawtchouk polynomials.

- All physically relevant orthogonal polynomials with a continuous spectrum that are compatible with SAQFT must have a sinusoidal asymptotic behavior. Specifically, in the limit as $n \to \infty$, we require that $p_n(z)$ takes the following form

$$p_n(z) \approx \frac{1}{n^\kappa \sqrt{\rho(z)}} \cos\left[n^\nu \varphi(z) + \delta(z)\right]\,, \tag{40}$$

  where $\kappa$ and $\nu$ are positive real parameters, $\varphi(z)$ is an entire function, and $\delta(z)$ is the phase shift. If $\nu \to 0$ then $n^\nu \to \ln(n)$. Only under this asymptotic condition, will the series (3a) produce oscillatory continuum states at the boundaries of space (see, for example, [19,27-30]). For a rigorous discussion about the connection between the asymptotics of such polynomials and scattering, one may consult [31-33] and references therein. Fortunately, all of the many known hypergeometric orthogonal polynomials that appear abundantly in the physics literature do meet this requirement. They include, but not limited to, the polynomials in the Askey scheme [26] such as the Wilson, continuous Hahn, continuous dual Hahn, Meixner-Pollaczek, Jacobi, Laguerre, Gegenbauer, Chebyshev, Hermite, etc.

- A more general differential equation satisfied by the complete set of functions $\{\phi_n(\vec{r})\}$ that maintains the tridiagonal algebraic structure of the theory is

$$\mathcal{D}\phi_n(\vec{r}) = \omega(\vec{r})\left[(\alpha_n - z)\phi_n(\vec{r}) + \beta_{n-1}\phi_{n-1}(\vec{r}) + \beta_n\phi_{n+1}(\vec{r})\right], \tag{41}$$

where $\mathcal{D}$ is the free wave operator (e.g., the Klein-Gordon, Dirac, Dirac-Coulomb, etc.) and $\omega(\vec{r})$ is an entire function that may vanish only at the boundaries. This equation should be considered a generalization of Eq. (4). For example, in Eq. (20), $\mathcal{D} = -\frac{d^2}{dx^2} + \lambda^2 W(x) + M^2 - E^2$ with $\omega(\vec{r}) = \lambda^2$, which is a generalization of the Klein-Gordon equation in 1+1 space-time where $W(x) \mapsto 0$.

- The linear momentum $\vec{k}$ (where $\vec{k}^2 = E^2 - M^2$) is an essential variable in the formulation of conventional QFT, which is usually written in the *k*-space representation. On the other hand, $\vec{k}$ does not appear explicitly in SAQFT. This is due to the tridiagonal structure of the fundamental differential equation for $\phi_n(\vec{r})$. In fact, using Eq. (4) or Eq. (20) in the Klein-Gordon wave equation gives $\vec{k}^2$ as one of the eigenvalues** of the following infinite tridiagonal symmetric matrix

$$\begin{pmatrix} \alpha_0 & \beta_0 & & & & & \\ \beta_0 & \alpha_1 & \beta_1 & & & & \\ & \beta_1 & \alpha_2 & \beta_2 & & & \\ & & \beta_2 & \alpha_3 & \beta_3 & & \\ & & & \times & \times & \times & \\ & & & & \times & \times & \times \\ & & & & & \times & \times \end{pmatrix}. \tag{42}$$

- For special systems where the continuous and discrete channels are totally independent, SAQFT formulation could be extended by choosing two distinct sets of spatial functions in the expansion of the continuous and discrete kernels $\psi(E,\vec{r})$ and $\psi_j(\vec{r})$. That is, we could still expand $\psi(E,\vec{r})$ in the same set $\{\phi_n(\vec{r})\}$ as given by (3a) but expand $\psi_j(\vec{r})$ in another *independent* set $\{\chi_n(\vec{r})\}$ by rewriting (3b) as

$$\psi_j(\vec{r}) = \sum_{n=0}^{\infty} g_n(E_j)\,\chi_n(\vec{r}) = g_0(E_j)\sum_{n=0}^{\infty} q_n(z_j)\,\chi_n(\vec{r}), \tag{43}$$

where $\{q_n(z_j)\}$ is a set of orthogonal spectral polynomials having *pure* discrete spectrum, whereas $\{p_n(z)\}$ has a *pure* continuous spectrum. Moreover, $\sum_{n=0}^{\infty}\chi_n(\vec{r})\,\bar{\chi}_n(\vec{r}') = \sum_{n=0}^{\infty}\bar{\chi}_n(\vec{r})\,\chi_n(\vec{r}') = \lambda^{-1}\delta^3(\vec{r}-\vec{r}')$ and $\langle\chi_n(\vec{r})|\bar{\chi}_m(\vec{r})\rangle = \langle\bar{\chi}_n(\vec{r})|\chi_m(\vec{r})\rangle = \lambda^{-1}\delta_{n,m}$. This allows for the flexibility of using two different set of spatial functions, one is suitable for the continuous channel and the other is more appropriate for the discrete channel (structure). The wave equation (4) or its generalization (41) for $\{\chi_n(\vec{r})\}$ maintains the same tridiagonal form but with different recursion coefficients associated with $\{q_n(z_j)\}$. Furthermore, $f_0^2(E)dE = \frac{1}{2}\rho(z)dz$ and $g_0^2(E_j) = \frac{1}{2}\lambda\xi(z_j)$ where we obtain

$$\left[\Psi(t,\vec{r}),\bar{\Psi}(t',\vec{r}')\right] = \lambda^{-1}\Delta(t-t',\vec{r}-\vec{r}') = \frac{1}{2}\sum_{n,m=0}^{\infty}\phi_n(\vec{r})\bar{\phi}_m(\vec{r}')\int_{\Omega} e^{-iE(t-t')}\rho(z)p_n(z)p_m(z)dz + \frac{1}{2}\sum_{n,m=0}^{\infty}\chi_n(\vec{r})\bar{\chi}_m(\vec{r}')\sum_{j=0}^{N} e^{-iE_j(t-t')}\xi(z_j)q_n(z_j)q_m(z_j) \tag{44}$$

giving $\Delta(0,\vec{r}-\vec{r}') = \delta^3(\vec{r}-\vec{r}')$. A remarkable example of such a system with an infinite but spatially confined structure is that for which $p_n(z)$ is the Meixner-Pollaczek polynomial $P_n^{\mu}(z,\theta)$ whereas $q_n(z_j)$ is the Meixner polynomial $M_n^{\mu}(z_j,\vartheta)$, [26]

$$P_n^{\mu}(z,\theta) = \sqrt{\frac{\Gamma(n+2\mu)}{\Gamma(n+1)\Gamma(2\mu)}}\, e^{in\theta}\, {}_2F_1\left(\begin{matrix}-n,\mu+iz\\ 2\mu\end{matrix}\middle|1-e^{-2i\theta}\right), \tag{45a}$$

$$M_n^{\mu}\left(z_j,\vartheta\right) = \sqrt{\frac{\Gamma(n+2\mu)}{\Gamma(n+1)\Gamma(2\mu)}}\, \vartheta^{n/2}\, {}_2F_1\left(\begin{matrix}-n,-j\\ 2\mu\end{matrix}\middle|1-\vartheta^{-1}\right), \tag{45b}$$

where $z_j = j+\mu$, $\mu>0$, $0<\theta<\pi$, $0<\vartheta<1$, and $j=0,1,2,...$. Another interesting example, but with a finite structure, is where $p_n(z)$ becomes the continuous dual Hahn polynomial $S_n^{\mu}(z;\sigma,\tau)$ with $(\mu,\sigma,\tau)>0$ and $q_n(z_j)$ is the dual Hahn polynomial $R_n^N(z_j;\sigma,\tau)$, [26]

$$R_n^N(z_j;\sigma,\tau) = \sqrt{\frac{(\sigma)_n(N-n+1)_n}{n!(N+\tau-n)_n}}\, {}_3F_2\left(\begin{matrix}-n,-j,j+\sigma+\tau-1\\ \sigma,-N\end{matrix}\middle|1\right), \tag{46}$$

where $z_j = \left(j+\frac{\sigma+\tau-1}{2}\right)^2$ and $n,j=0,1,..,N$.

---

** Let $\Sigma$ be the matrix whose elements are $\langle\phi_n|\phi_m\rangle$, then $\vec{k}^2$ is a generalized eigenvalue in the matrix equation $Q|..\rangle = \vec{k}^2\Sigma|..\rangle$ where *Q* is the tridiagonal symmetric matrix (42). If $\bar{\phi}_n = \phi_n$, then $\Sigma_{n,m} = \delta_{n,m}$ and $Q|..\rangle = \vec{k}^2|..\rangle$.

- By incorporating hadron structure, we conjecture that SAQFT could be used to develop a QFT for hadrons (baryons and mesons) without the need for color charges or nonabelian quantum fields. That is, an alternative to QCD for nuclear physics applications. In that case, the spectral polynomial $p_n(z)$ associated with a three-quark baryon could be taken as the Wilson polynomial $W_n^{\mu}(z;a,b,c)$ [26] with $-3<\mu<-2$ whereas the spectral polynomial associated with a two-quark meson would be the continuous dual Hahn polynomial $S_n^{\mu}(z;a,b)$ with $-2<\mu<-1$. Each of the parameters $\{a,b,c\}$ can assume one of 6 values corresponding to one of the six flavors of quarks or their conjugate values for the six anti-quarks.
- In Section 4, while calculating the scattering amplitudes to second order using the Feynman diagrams for the given model, we obtained finite values for the loop integrals. We believe that all such loop integrals at higher orders of perturbation for that model are finite because the fundamental SAQFT integrals of Eq. (30) and other higher order loop integrals like the following:

$$\int_{\Omega}\rho(x)\rho(x\circledast y)p_n^2(x)p_m^2(x\circledast y)dxdy, \tag{47a}$$

$$\int_\Omega \rho(x)\rho(x\circledast a)\rho(y\circledast a)\,p_n^2(x)\,p_m^2(x\circledast a)\,p_k^2(y\circledast a)\,dxdy\,, \tag{47b}$$

$$\int_\Omega \rho(x)\rho(x\circledast y)\rho(y\circledast a)\,p_n^2(x)\,p_m^2(x\circledast y)\,p_k^2(y\circledast a)\,dxdy\,, \tag{47c}$$

$$\int_\Omega \rho(x)\rho(x\circledast y)\rho(x\circledast z)\,p_n^2(x)\,p_m^2(x\circledast y)\,p_k^2(x\circledast z)\,dxdydz\,, \tag{47d}$$

$$\int_\Omega \rho(x)\rho(x\circledast y)\rho(y\circledast z)\,p_n^2(x)\,p_m^2(x\circledast y)\,p_k^2(y\circledast z)\,dxdydz\,, \tag{47e}$$

---

---

$$\zeta_{n,m,..,k}(a,b,..,c)=\int_\Omega \rho(x)\rho(y\circledast a)...\rho(z\circledast c)\,p_n^2(x)\,p_m^2(y\circledast a)...p_k^2(z\circledast c)\,dxdy...dz\,, \tag{48}$$

are not just finite but fall within the range of values $[0,+1]$. In these fundamental SAQFT integrals, the number of indices $\{n,m,..,k\}$ (spectral polynomial degrees) represents the number of vertices whereas the number of integrations over the spectral parameters $\{x,y,..,z\}$ stands for the number of loops in the diagrams (perturbation order). We have verified our claim numerically up to three loops and for several choices of spectral polynomials in the model of Section 4. If this could be confirmed for all perturbation orders then it means that SAQFT for this interaction model is a finite theory doing away with renormalization. It remains to be seen whether this SAQFT finiteness property extends to other physically relevant models.

- The tridiagonal representation approach (TRA) is an algebraic method for solving linear ordinary differential equations of the second order [34,35]. It has been used successfully in the solution of the wave equation in quantum mechanics resulting in a larger class of exactly solvable problems (see, for example, [36] and references therein). The *free* sector of SAQFT presented here, which is linear, could be viewed as an application of the TRA in QFT.

## Appendix A: The continuous dual Han polynomial

The orthonormal version of the three-parameter continuous dual Han polynomial reads as follows

$$S_n^\mu(x^2;\sigma,\tau)=\sqrt{\frac{(\mu+\sigma)_n(\mu+\tau)_n}{n!\,(\sigma+\tau)_n}}\;{}_3F_2\left(\left.\begin{matrix}-n,\mu+\mathrm{i}x,\mu-\mathrm{i}x\\ \mu+\sigma,\mu+\tau\end{matrix}\right|1\right), \tag{A1}$$

where ${}_3F_2\left(\left.\begin{matrix}a,b,c\\ d,e\end{matrix}\right|x\right)=\sum_{n=0}^{\infty}\frac{(a)_n(b)_n(c)_n}{(d)_n(e)_n}\frac{x^n}{n!}$ is the generalized hypergeometric series and $(a)_n=a(a+1)(a+2)...(a+n-1)=\frac{\Gamma(n+a)}{\Gamma(a)}$ is the Pochhammer symbol (shifted factorial). Moreover, if $\mu>0$ and $\mathrm{Re}(\sigma,\tau)>0$ with non-real parameters occurring in conjugate pairs then this is a

polynomial in $x^2$ with a pure continuous spectrum. It is orthogonal with respect to the measure $\rho^\mu(x;\sigma,\tau)\,dx$ whose normalized form reads as follows

$$\rho^\mu(x;\sigma,\tau)=\frac{1}{2\pi}\frac{\left|\Gamma(\mu+\mathrm{i}x)\Gamma(\sigma+\mathrm{i}x)\Gamma(\tau+\mathrm{i}x)/\Gamma(2\mathrm{i}x)\right|^2}{\Gamma(\mu+\sigma)\Gamma(\mu+\tau)\Gamma(\sigma+\tau)}. \tag{A2}$$

That is, $\int_0^\infty S_n^\mu(x^2;\sigma,\tau)S_m^\mu(x^2;\sigma,\tau)\rho^\mu(x;\sigma,\tau)\,dx=\delta_{nm}$. However, if the parameters are such that $\mu<0$ and $\sigma+\mu$, $\tau+\mu$ are positive or a pair of complex conjugates with positive real parts, then the polynomial will have a continuous spectrum as well as a finite size discrete spectrum and the polynomial satisfies the following generalized orthogonality relation

$$\int_0^\infty \rho^\mu(x;\sigma,\tau)\,S_n^\mu(x^2;\sigma,\tau)\,S_m^\mu(x^2;\sigma,\tau)\,dx+\sum_{j=0}^{N}\xi^\mu(x_j;\sigma,\tau)\,S_n^\mu(x_j^2;\sigma,\tau)\,S_m^\mu(x_j^2;\sigma,\tau)=\delta_{n,m}. \tag{A3}$$

where $N$ is the largest integer less than $-\mu$ and

$$\xi^\mu(x_j;\sigma,\tau)=2(-1)^j\,\frac{(-\mu-j)}{j!}\,\frac{(\mu+\sigma)_j\,(\mu+\tau)_j\,(2\mu)_j}{(\mu-\sigma+1)_j\,(\mu-\tau+1)_j}\,\frac{\Gamma(\sigma-\mu)\Gamma(\tau-\mu)}{\Gamma(\sigma+\tau)\Gamma(1-2\mu)}. \tag{A4}$$

The large degree asymptotics ($n\to\infty$) of $S_n^\mu(x^2;\sigma,\tau)$, which could be found in the Appendix of Ref. [29], vanishes if $\mu+\mathrm{i}x=-j$, where $j=0,1,2,..,N$. Thus, the spectrum formula associated with this polynomial is

$$x_j^2=-(j+\mu)^2. \tag{A5}$$

Moreover, these polynomials satisfy the following symmetric three-term recursion relation

$$\begin{aligned}x^2S_n^\mu(x^2;\sigma,\tau)&=\left[(n+\mu+\sigma)(n+\mu+\tau)+n(n+\sigma+\tau-1)-\mu^2\right]S_n^\mu(x^2;\sigma,\tau)\\&-\sqrt{n(n+\sigma+\tau-1)(n+\mu+\sigma-1)(n+\mu+\tau-1)}\,S_{n-1}^\mu(x^2;\sigma,\tau)\\&-\sqrt{(n+1)(n+\sigma+\tau)(n+\mu+\sigma)(n+\mu+\tau)}\,S_{n+1}^\mu(x^2;\sigma,\tau)\end{aligned} \tag{A6}$$

## Appendix B: Dirac spinor in SAQFT

We provide in this Appendix a brief description of the four-component quantum field $\Psi^{\uparrow\downarrow}(t,\vec{r})$ representing the Dirac spinor with structure in 3+1 Minkowski space-time. If we adopt the conventional representation of the $4\times4$ Dirac gamma matrices as $\gamma^0=\begin{pmatrix}\mathrm{I}&0\\0&-\mathrm{I}\end{pmatrix}$ and $\vec{\gamma}=\begin{pmatrix}0&\vec{\sigma}\\-\vec{\sigma}&0\end{pmatrix}$ where I is the $2\times2$ unit matrix and $\{\vec{\sigma}\}$ are the three Pauli spin matrices, then the free Dirac equation for $\Psi^{\uparrow\downarrow}(t,\vec{r})$ reads $\sum_{\mu=0}^{\mu=3}\left(\mathrm{i}\gamma^\mu\partial_\mu-M\right)\Psi^{\uparrow\downarrow}(t,\vec{r})=0$. Multiplying the equation from left by $\gamma^0$, we obtain the following $4\times4$ matrix equation

$$\begin{pmatrix}(\mathrm{i}\partial_t-M)\mathrm{I} & \mathrm{i}\vec{\sigma}\cdot\vec{\nabla}\\ \mathrm{i}\vec{\sigma}\cdot\vec{\nabla} & (\mathrm{i}\partial_t+M)\mathrm{I}\end{pmatrix}\begin{pmatrix}\Psi_+^{\uparrow\downarrow}(t,\vec{r})\\ \Psi_-^{\uparrow\downarrow}(t,\vec{r})\end{pmatrix}=0. \tag{B1}$$

The two components of the quantum spinor field are written as

$$\Psi_{\pm}^{\uparrow}(t,\vec{r}) = \int_{\Omega} e^{-\mathrm{i}Et}\psi_{\pm}^{\uparrow}(E,\vec{r})\, b_{\uparrow}(E)\, dE + \sum_{j=0}^{N} e^{-\mathrm{i}E_j t}\psi_{\pm}^{\uparrow,j}(\vec{r})\, b_{\uparrow}^{j}, \tag{B2a}$$

$$\Psi_{\pm}^{\downarrow}(t,\vec{r}) = \int_{\Omega} e^{-\mathrm{i}Et}\psi_{\pm}^{\downarrow}(E,\vec{r})\, b_{\downarrow}(E)\, dE + \sum_{j=0}^{N} e^{-\mathrm{i}E_j t}\psi_{\pm}^{\downarrow,j}(\vec{r})\, b_{\downarrow}^{j}, \tag{B2b}$$

where $\Omega = \left\{E : |E| \geq M\right\}$ and $\left\{|E_j| < M\right\}_{j=0}^{N}$. The creation and annihilation operators satisfy the following anti-commutation relations

$$\left\{b_s(E), b_{s'}^{\dagger}(E')\right\} = b_s(E)\, b_{s'}^{\dagger}(E') + b_{s'}^{\dagger}(E')\, b_s(E) = \delta_{s,s'}\,\delta(E-E'), \quad \left\{b_s^{i}, (b_{s'}^{j})^{\dagger}\right\} = \lambda^{-1}\delta_{s,s'}\,\delta^{i,j}. \tag{B3}$$

where $s$ and $s'$ stand for the $\uparrow\downarrow$ spins.[††] All other anti-commutators among $b_s(E)$, $b_s^{\dagger}(E)$, $b_s^{j}$, and $(b_s^{j})^{\dagger}$ vanish. The Fourier energy components are written as pointwise convergent series in terms of the two-component complete set of spatial functions $\left\{\phi_n^{\pm}(\vec{r})\right\}$:

$$\psi_{\pm}^{s}(E,\vec{r}) = \sum_{n=0}^{\infty} f_n^{s}(E)\,\phi_n^{\pm}(\vec{r}) = f_0^{s}(E)\sum_{n=0}^{\infty} p_n^{s}(z)\,\phi_n^{\pm}(\vec{r}), \tag{B4a}$$

$$\psi_{\pm}^{s,j}(\vec{r}) = \sum_{n=0}^{\infty} g_n^{s}(E_j)\,\phi_n^{\pm}(\vec{r}) = g_0^{s}(E_j)\sum_{n=0}^{\infty} p_n^{s}(z_j)\,\phi_n^{\pm}(\vec{r}). \tag{B4b}$$

The spectral parameter $z$ is to be determined and $\left\{\phi_n^{\pm}(\vec{r})\right\}$ must satisfy the following coupled differential relations

$$-\mathrm{i}\vec{\sigma}\cdot\vec{\nabla}\phi_n^{-}(\vec{r}) = c_n\,\phi_n^{+}(\vec{r}) + d_n\,\phi_{n-1}^{+}(\vec{r}), \tag{B5a}$$

$$-\mathrm{i}\vec{\sigma}\cdot\vec{\nabla}\phi_n^{+}(\vec{r}) = c_n\,\phi_n^{-}(\vec{r}) + d_{n+1}\,\phi_{n+1}^{-}(\vec{r}), \tag{B5b}$$

where $\left\{c_n, d_n\right\}$ are non-zero constants. The first (second) relation is referred to as the forward (backward) shift operator. Substituting (B4) into (B2) and using (B5), the coupled free Dirac equation (B1) becomes the following set of two difference relations

$$\left(E+M\right) p_n^{s}(z) = c_n\, p_n^{s}(z) + d_n\, p_{n-1}^{s}(z), \quad \text{for } E > -M. \tag{B6a}$$

$$\left(E-M\right) p_n^{s}(z) = c_n\, p_n^{s}(z) + d_{n+1}\, p_{n+1}^{s}(z), \quad \text{for } E < M. \tag{B6b}$$

Multiplying both sides of Eq. (B6a) by $E-M$ for $E<M$ and using Eq. (B6b), we obtain the following algebraic relation

$$\left(E^2 - M^2\right) p_n^{s}(z) = \left(c_n^2 + d_n^2\right) p_n^{s}(z) + \left(c_{n-1}\, d_n\right) p_{n-1}^{s}(z) + \left(c_n\, d_{n+1}\right) p_{n+1}^{s}(z). \tag{B7a}$$

[††] One may consider an alternative definition of the anti-commutation relation for the discrete creation/annihilation operators as $\left\{b_s^{i}, (b_{s'}^{j})^{\dagger}\right\} = \delta_{s,s'}\,\delta^{i,j}\big/(E_j + M)$. In this case the formula $[g_0^{s}(E_j)]^2 = \lambda\xi^{s}(z_j)$ written below Eq. (B11) must be changed to read $[g_0^{s}(E_j)]^2 = (E_j + M)\xi^{s}(z_j)$.

On the other hand, if we multiply both sides of Eq. (B6b) by $E+M$ for $E>-M$ and use Eq. (B6a), we obtain

$$\left(E^2-M^2\right)p_n^s(z)=\left(c_n^2+d_{n+1}^2\right)p_n^s(z)+\left(c_n\,d_n\right)p_{n-1}^s(z)+\left(c_{n+1}\,d_{n+1}\right)p_{n+1}^s(z)\,. \qquad \text{(B7b)}$$

One of these two relations is associated with spin up and the other with spin down. Both are identical to the three-term recursion relation (5) where we adopt the following parameter assignments

$$z=E^2-M^2\,,\qquad \alpha_n=c_n^2+d_n^2:=\alpha_n^\uparrow\,,\qquad \beta_n=c_n\,d_{n+1}:=\beta_n^\uparrow\,. \qquad \text{(B8a)}$$

$$z=E^2-M^2\,,\qquad \alpha_n=c_n^2+d_{n+1}^2:=\alpha_n^\downarrow\,,\qquad \beta_n=c_{n+1}\,d_{n+1}:=\beta_n^\downarrow\,. \qquad \text{(B8b)}$$

The recursion coefficients of the spectral polynomials $p_n^{\uparrow\downarrow}(z)$ can be transformed into each other by the parameter map $c_n\mapsto d_{n+1}$ and $d_n\mapsto c_n$. Therefore, $\left\{p_n^s(z)\right\}$ is a sequence of orthogonal spectral polynomials in $z$ that satisfy the symmetric three-term recursion relation (5) with $\left\{\alpha_n,\beta_n\right\}\mapsto\left\{\alpha_n^s,\beta_n^s\right\}$ shown in (B8). The initial values are $p_0^s(z)=1$ and $p_1^s(z)=(z-\alpha_0^s)/\beta_0^s$. Moreover, $\left\{p_n^s(z)\right\}$ are also required to fulfill the general orthogonality relation (6) but with $\left\{\rho(z),\xi(z_j)\right\}\mapsto\left\{\rho^s(z),\xi^s(z_j)\right\}$.

The conjugate quantum field $\bar\Psi^{\uparrow\downarrow}(t,\vec r)$ is obtained from (B2) by the maps $\Psi_\pm^{\uparrow\downarrow}\mapsto(\Psi_\pm^{\uparrow\downarrow})^\dagger$ and $\phi_n^\pm\mapsto\bar\phi_n^\pm$ where

$$\left\langle\phi_n^r(\vec r)\middle|\bar\phi_m^{r'}(\vec r)\right\rangle=\left\langle\bar\phi_n^r(\vec r)\middle|\phi_m^{r'}(\vec r)\right\rangle=\delta_{r,r'}\,\delta_{n,m}\,, \qquad \text{(B9a)}$$

$$\sum_{n=0}^\infty\phi_n^r(\vec r)\bar\phi_n^{r'}(\vec r')=\sum_{n=0}^\infty\bar\phi_n^r(\vec r)\phi_n^{r'}(\vec r')=\delta_{r,r'}\,\delta^3(\vec r-\vec r')\,, \qquad \text{(B9b)}$$

where $r$ and $r'$ stand for the four spinor components $1,2,3,4$ with $\phi_n^+=\begin{pmatrix}\phi_n^1\\\phi_n^2\end{pmatrix}$ and $\phi_n^-=\begin{pmatrix}\phi_n^3\\\phi_n^4\end{pmatrix}$. One should note the absence of the parameter $\lambda$ from these formulas in contrast with the corresponding formulas (8a) and (8b) in scalar SAQFT. Thus, in $n+1$ space-time, the length scale of $\phi_n(\vec r)$ and $\bar\phi_n(\vec r)$ is $-n/2$. If we adopt the conventional notation that $\bar\phi_n=\chi_n^\dagger\gamma^0$ [i.e., $\bar\phi_n^\pm=\pm(\chi_n^\pm)^\dagger$] then we obtain the orthogonality $\sum_{r=1}^4\left\langle\phi_n^r(\vec r)\middle|\chi_m^r(\vec r)^*\right\rangle=\sum_{r=1}^4\left\langle\chi_n^r(\vec r)^*\middle|\phi_m^r(\vec r)\right\rangle=0$ and we can rewrite (B9) as

$$\left\langle\phi_n(\vec r)\middle|\chi_m^\dagger(\vec r)\right\rangle=\left\langle\phi_m(\vec r)\middle|\chi_n^\dagger(\vec r)\right\rangle=\gamma^0\,\delta_{n,m}\,. \qquad \text{(B9a)'}$$

$$\sum_{n=0}^\infty\phi_n(\vec r)\,\chi_n^\dagger(\vec r')=\sum_{n=0}^\infty\phi_n(\vec r')\,\chi_n^\dagger(\vec r)=\gamma^0\delta^3(\vec r-\vec r')\,, \qquad \text{(B9b)'}$$

Moreover, we can write $\bar\Psi^{\uparrow\downarrow}(t,\vec r)$ as follows

$$\bar\Psi_\pm^s(t,\vec r)=\int_\Omega e^{iEt}\bar\psi_\pm^s(E,\vec r)b_s^\dagger(E)dE+\sum_{j=0}^N e^{iE_jt}\bar\psi_\pm^{s,j}(\vec r)(b_s^j)^\dagger\,. \qquad \text{(B10)}$$

where the Fourier energy components $\bar{\psi}^{s}_{\pm}(E,\vec{r})$ and $\bar{\psi}^{s,j}_{\pm}(\vec{r})$ are identical to (B4) but with $\phi^{\pm}_{n}(\vec{r}) \mapsto \bar{\phi}^{\pm}_{n}(\vec{r})$. Using the anti-commutators (B3), we can write

$$\left\{\Psi^{s}_{r}(t,\vec{r}),\bar{\Psi}^{s'}_{r'}(t',\vec{r}')\right\} = \delta_{s,s'} \sum_{n,m=0}^{\infty} \phi^{r}_{n}(\vec{r})\bar{\phi}^{r'}_{m}(\vec{r}') \times\left[\int_{\Omega} e^{-\mathrm{i}E(t-t')}\rho^{s}(z)p^{s}_{n}(z)p^{s}_{m}(z)dz + \sum_{j=0}^{N} e^{-\mathrm{i}E_{j}(t-t')}\xi^{s}(z_{j})p^{s}_{n}(z_{j})p^{s}_{m}(z_{j})\right] \tag{B11}$$

where we wrote $[f^{s}_{0}(E)]^{2}dE = \rho^{s}(z)dz$ with $dz/dE > 0$ for $E\in\Omega$ and $[g^{s}_{0}(E_{j})]^{2} = \lambda\xi^{s}(z_{j})$ for $\left|E_{j}\right| < M$. As in conventional QFT, this defines the singular distribution $\Delta_{r,r'}(t-t',\vec{r}-\vec{r}')$ by

$$\left\{\Psi^{s}_{r}(t,\vec{r}),\bar{\Psi}^{s'}_{r'}(t',\vec{r}')\right\} = \delta_{s,s'}\,\Delta_{r,r'}(t-t',\vec{r}-\vec{r}'). \tag{B12}$$

Using the orthogonality (6) of the spectral polynomials $\left\{p^{s}_{n}(z)\right\}$ and the completeness (B9b) of the set $\left\{\phi^{r}_{n}(\vec{r})\right\}$, Eq. (B11) with $t = t'$ becomes

$$\left\{\Psi^{s}_{r}(t,\vec{r}),\bar{\Psi}^{s'}_{r'}(t,\vec{r}')\right\} = \delta_{s,s'}\,\delta_{r,r'}\,\delta^{3}(\vec{r}-\vec{r}'), \tag{B13}$$

Equations (B12) and (B13) give $\Delta_{r,r'}(0,\vec{r}-\vec{r}') = \delta_{r,r'}\,\delta^{3}(\vec{r}-\vec{r}')$. Moreover, it is straightforward to write

$$\left\{\Psi^{s}_{r}(t,\vec{r}),\Psi^{s'}_{r'}(t,\vec{r}')\right\} = \left\{\bar{\Psi}^{s}_{r}(t,\vec{r}),\bar{\Psi}^{s'}_{r'}(t,\vec{r}')\right\} = 0. \tag{B14}$$

Therefore, the canonical conjugate to the spinor quantum field $\Psi^{s}_{r}(t,\vec{r})$ becomes $\Pi^{s}_{r}(t,\vec{r}) = \mathrm{i}\bar{\Psi}^{s}_{r}(t,\vec{r})$. Additionally, and as done above for the scalar particle, we can define the positive-energy non-elementary spinor particle by the quantum field

$$\mathcal{X}_{\pm}(t,\vec{r}) = \frac{1}{\sqrt{2}}\left[\mathbf{\Psi}^{\uparrow}_{\pm}(t,\vec{r}) + \mathbf{\Psi}^{\downarrow}_{\pm}(t,\vec{r})\right], \tag{B15a}$$

where $\mathbf{\Psi}^{s}_{\pm}(t,\vec{r}) = \Psi^{s}_{\pm}(t,\vec{r}) \pm \hat{\Psi}^{s}_{\pm}(t,\vec{r})^{\dagger}$ and $\hat{\Psi}^{s}_{r}(t,\vec{r})$ is identical to $\Psi^{s}_{r}(t,\vec{r})$ except that $b_{s}(E) \mapsto \hat{b}_{s}(E)$ and $b^{j}_{s} \mapsto \hat{b}^{j}_{s}$, which are independent annihilation fermion operators associated with the second independent solution of the Dirac equation. The $\pm$ sign to the left of $\hat{\Psi}^{s}_{\pm}(t,\vec{r})^{\dagger}$ comes from multiplication (on the right) of the 4-component spinor $(\hat{\Psi}^{s})^{\dagger}$ by the gamma matrix $\gamma^{0}$. On the other hand, the corresponding anti-particle is represented by the negative-energy quantum field

$$\bar{\mathcal{X}}_{\pm}(t,\vec{r}) = \frac{1}{\sqrt{2}}\left[\bar{\mathbf{\Psi}}^{\uparrow}_{\pm}(t,\vec{r}) + \bar{\mathbf{\Psi}}^{\downarrow}_{\pm}(t,\vec{r})\right]. \tag{B15b}$$

The Feynman propagator is obtained as follows

$$\mathrm{i}\Delta^F_{r',r}(t'-t,\vec{r}'-\vec{r}) = \langle 0|T\left(\bar{\mathcal{X}}_{r'}(t',\vec{r}'),\mathcal{X}_r(t,\vec{r})\right)|0\rangle = \langle 0|\bar{\mathcal{X}}_{r'}(t',\vec{r}')\mathcal{X}_r(t,\vec{r})|0\rangle\theta(t'-t) - \langle 0|\mathcal{X}_r(t,\vec{r})\bar{\mathcal{X}}_{r'}(t',\vec{r}')|0\rangle\theta(t-t') \tag{B16}$$

Note the minus sign in the middle of the second line of this propagator due to anti-commutation of the field operators. Also, note that for real (neutral) spinors (e.g., the neutrinos), $\bar{\phi}_n = \phi_n^\dagger \gamma^0$.

## Appendix C: Massless vector field in SAQFT

In this Appendix, we give a brief description of the massless vector field in SAQFT. The particle associated with this quantum field (e.g., the photon) has two degrees of freedom corresponding to the two states of polarizations that transverse its propagation, which we designate by the superscripts $\rightleftarrows$. It is constructed using the solution space of the free wave equation: $\Box A^{\rightleftarrows}_\mu(t,\vec{r}) = \left(\partial_t^2 - \vec{\nabla}^2\right)A^{\rightleftarrows}_\mu(t,\vec{r}) = 0$. The space is reduced from four degrees of freedom carried by $A^s_\mu$ to two by:

(1) Removing the gauge modes of the form $A^s_\mu = \partial_\mu \Phi^s$ where $\Phi^s$ is a dimensionless space-time scalar function, and

(2) Imposing the Lorenz invariant Landau gauge fixing condition $\sum_{\mu=0}^{\mu=3}\partial^\mu A^s_\mu = 0$,

where *s* stands for the polarizations $\rightleftarrows$.

As explained in the main text following Eq. (1), massless particles in SAQFT are necessarily structureless. Hence, the summation part in the energy expansion of the quantum field (the particle structure) is absent and we write

$$A^{\rightleftarrows}_\mu(t,\vec{r}) = \int_\Omega e^{-\mathrm{i}Et}\psi^{\rightleftarrows}_\mu(E,\vec{r})\,a_{\rightleftarrows}(E)\,dE\,, \tag{C1}$$

where $\Omega = \{E \in \mathbb{R}\}$ and $\left[a_s(E), a^\dagger_{s'}(E')\right] = \delta_{s,s'}\,\delta(E-E')$. Moreover,

$$\psi^s_\mu(E,\vec{r}) = \sum_{n=0}^{\infty} f^s_n(E)\phi^\mu_n(\vec{r}) = f^s_0(E)\sum_{n=0}^{\infty} p^s_n(z)\phi^\mu_n(\vec{r})\,, \tag{C2}$$

where $\left\{f^s_n(E)\right\}$ are real and $\vec{\phi}_n(\vec{r}) \neq \vec{\nabla}\varphi_n(\vec{r})$. The Landau gauge fixing, $\partial^\mu A^s_\mu = 0$, becomes $E\phi^0_n(\vec{r}) = -\mathrm{i}\vec{\nabla}\cdot\vec{\phi}_n(\vec{r})$ and we require that:

$$\text{For } E>0: \qquad -\mathrm{i}\vec{\nabla}\cdot\vec{\phi}_n(\vec{r}) = c_n\phi^0_n(\vec{r}) + d_n\phi^0_{n-1}(\vec{r})\,, \tag{C3a}$$

$$\text{For } E<0: \qquad -\mathrm{i}\vec{\nabla}\cdot\vec{\phi}_n(\vec{r}) = c_n\phi^0_n(\vec{r}) + d_{n+1}\phi^0_{n+1}(\vec{r})\,, \tag{C3b}$$

where $\{c_n, d_n\}$ are non-zero constant parameters. Substituting (C3) in the Landau condition $\partial^\mu A^s_\mu = 0$, we obtain:

$$\text{For } E>0: \qquad E\,p^s_n(z) = c_n\,p^s_n(z) + d_{n+1}\,p^s_{n+1}(z)\,. \tag{C4a}$$

For $E<0$: $$E\,p_n^s(z)=c_n\,p_n^s(z)+d_n\,p_{n-1}^s(z)\,. \tag{C4b}$$

For $E<0$, we multiply both sides of (C4a) by *E* and use (C4b) giving

$$E^2 p_n^s(z)=\alpha_n^s p_n^s(z)+\beta_{n-1}^s p_{n-1}^s(z)+\beta_n^s p_{n+1}^s(z)\,, \tag{C5}$$

making $z=E^2$, $\alpha_n^s=c_n^2+d_{n+1}^2$, and $\beta_n^s=c_{n+1}\,d_{n+1}$. On the other hand, for $E>0$, we multiply both sides of (C4b) by *E* and use (C4a) giving the same three-term recursion relation (C5) but with $\alpha_n^s=c_n^2+d_n^2$ and $\beta_n^s=c_n\,d_{n+1}$. We associate the latter recursion coefficients $\left\{\alpha_n^s,\beta_n^s\right\}$ with $p_n^{\rightarrow}(z)$ and the former with $p_n^{\leftarrow}(z)$. Moreover, the free wave equation $\left(\partial_t^2-\vec{\nabla}^2\right)A_\mu^s(t,\vec{r})\;=0$ becomes the three-term recursion relation (C5) dictating that

$$-\vec{\nabla}^2\phi_n^\mu(\vec{r})=\alpha_n^s\phi_n^\mu(\vec{r})+\beta_{n-1}^s\phi_{n-1}^\mu(\vec{r})+\beta_n^s\phi_{n+1}^\mu(\vec{r})\,. \tag{C6}$$

In addition to the recursion relation (C5), the spectral polynomials $\left\{p_n^s(z)\right\}$ satisfy the following continuous orthogonality relation

$$\int_\Omega \rho^s(z)\,p_n^s(z)\,p_m^s(z)\,dz=\delta_{n,m}\,, \tag{C7}$$

where the positive definite weight function is written as $\rho^s(z)dz=[\,f_0^s(E)]^2dE$ with $\frac{dz}{dE}>0$.

The conjugate quantum field $\bar{A}_\mu^{\rightleftarrows}(t,\vec{r})$ is obtained from (C1) and (C2) by complex conjugation and $\phi_n^\mu(\vec{r})\mapsto\bar{\phi}_n^\mu(\vec{r})$ where

$$\left\langle\phi_n^\mu(\vec{r})\middle|\bar{\phi}_m^\nu(\vec{r})\right\rangle=\left\langle\bar{\phi}_n^\mu(\vec{r})\middle|\phi_m^\nu(\vec{r})\right\rangle=\lambda^{-1}\delta_{\mu\nu}\,\delta_{n,m}\,, \tag{C8a}$$

$$\sum_{n=0}^\infty\phi_n^\mu(\vec{r})\,\bar{\phi}_n^\nu(\vec{r}\,')=\sum_{n=0}^\infty\bar{\phi}_n^\mu(\vec{r})\,\phi_n^\nu(\vec{r}\,')=\lambda^{-1}\delta_{\mu\nu}\,\delta^3(\vec{r}-\vec{r}\,')\,. \tag{C8b}$$

Therefore, we can write $\bar{A}_\mu^{\rightleftarrows}(t,\vec{r})=\int_\Omega e^{\mathrm{i}Et}\bar{\psi}_\mu^{\rightleftarrows}(E,\vec{r})\,a_{\rightleftarrows}^\dagger(E)\,dE$ and with the use of the above properties, we obtain the equal time commutation relation of the field operators as follows

$$\left[A_\mu^s(t,\vec{r}),\bar{A}_\nu^{s'}(t,\vec{r}\,')\right]=\lambda^{-1}\delta_{s,s'}\,\delta_{\mu\nu}\,\delta^3(\vec{r}-\vec{r}\,')\,. \tag{C9}$$

However, in general, we write $\left[A_\mu^s(t,\vec{r}),\bar{A}_\nu^{s'}(t',\vec{r}\,')\right]=\lambda^{-1}\delta_{s,s'}\Delta_{\mu,\nu}(t-t',\vec{r}-\vec{r}\,')$ where

$$\Delta_{\mu,\nu}(t-t',\vec{r}-\vec{r}\,')=\lambda\sum_{n,m=0}^\infty\phi_n^\mu(\vec{r})\,\bar{\phi}_m^\nu(\vec{r}\,')\int_\Omega e^{-\mathrm{i}E(t-t')}\rho^s(z)\,p_n^s(z)\,p_m^s(z)\,dz\,. \tag{C10}$$

Thus, $\Delta_{\mu,\nu}(0,\vec{r}-\vec{r}\,')=\delta_{\mu\nu}\,\delta^3(\vec{r}-\vec{r}\,')$. The real (neutral) massless particle (e.g., the photon) is represented by the quantum field $\mathcal{A}_\mu(t,\vec{r})=\frac{1}{\sqrt{2}}\left[A_\mu(t,\vec{r})+\bar{A}_\mu(t,\vec{r})\right]$ where $A_\mu(t,\vec{r}):=A_\mu^{\rightarrow}(t,\vec{r})+A_\mu^{\leftarrow}(t,\vec{r})$ and $\bar{\phi}_n^\mu(\vec{r})=\phi_n^\mu(\vec{r})^*$. However, the complex (charged) massless particle is represented by the positive-energy quantum field $\mathcal{A}_\mu(t,\vec{r}):=\frac{1}{\sqrt{2}}\left[\mathbf{A}_\mu^{\rightarrow}(t,\vec{r})+\mathbf{A}_\mu^{\leftarrow}(t,\vec{r})\right]$ where $\mathbf{A}_\mu^s(t,\vec{r}):=A_\mu^s(t,\vec{r})+\hat{A}_\mu^s(t,\vec{r})^\dagger$ with $\hat{A}_\mu^s(t,\vec{r})$ being identical to $A_\mu^s(t,\vec{r})$ except that $a_s(E)\mapsto$ $\hat{a}_s(E)$, which is an independent annihilation operator associated with the second independent solution of the massless Klein-Gordon equation. The anti-particle is represented by the negative-energy quantum field $\bar{\mathcal{A}}_\mu(t,\vec{r}) := \frac{1}{\sqrt{2}}\left[\bar{\mathbf{A}}_\mu^{\rightarrow}(t,\vec{r}) + \bar{\mathbf{A}}_\mu^{\leftarrow}(t,\vec{r})\right]$. The associated Feynman propagator is defined in the usual way using these positive and negative energy quantum fields.

A practical first study in SAQFT would be to reproduce some of the well-known results in QED where the photon is represented by the quantum field $\mathcal{A}_\mu(t,\vec{r})$ and the associated spectral polynomial could be taken as either the Hermite polynomial $H_n(k/\lambda)$ or the Gegenbauer polynomial $C_n^\nu(\cos\theta)$ where $\cos\theta = \left(z - \frac{1}{4}\lambda^2\right)/\left(z + \frac{1}{4}\lambda^2\right)$. The electron is represented by the structureless quantum field $\mathcal{X}(t,\vec{r})$ with the associated spectral polynomial taken as either the Laguerre polynomial $L_n^\nu(z)$ or the Meixner-Pollaczek polynomial $P_n^\nu(\cos\vartheta,\vartheta)$ where $\cos\vartheta = \left(z - \frac{1}{2}\lambda^2\right)/\left(z + \frac{1}{2}\lambda^2\right)$. The interaction Lagrangian is $\mathscr{L}_I = -e \otimes \sum_{\mu=0}^{3} \mathcal{A}_\mu(t,\vec{r})[\bar{\mathcal{X}}(t,\vec{r})\gamma^\mu \mathcal{X}(t,\vec{r})]$, which is cubic in the field operators like the model studied in Section 4.

With $A_\mu^{\rightleftarrows}(t,\vec{r})$ representing the massless gauge vector field that mediates interaction, we can formulate the non-abelian version of SAQFT by constructing the following quantum field matrices

$$\mathbb{A}_\mu^{\rightleftarrows}(t,\vec{r}) := \sum_{k=1}^{K^2-1} [A_\mu^{\rightleftarrows}(t,\vec{r})]^k T_k \,, \tag{C11}$$

where $\{T_k\}$ are the $K^2 - 1$ Hermitian generators of SU($K$) with a Lie algebra and normalization

$$\left[T_i, T_j\right] = \mathrm{i}\sum_{k=1}^{K^2-1} C_{i,j}^k T_k \,, \qquad \mathrm{Tr}\left(T_i\, T_j\right) = \frac{1}{2}\delta_{i,j} \,. \tag{C12}$$

An example of an interaction Lagrangian in a non-abelian SAQFT involving spinors and the gauge vector field could read:

$$\mathscr{L}_I = g \otimes \sum_{k=1}^{K^2-1} \sum_{\mu=0}^{3} [\mathcal{A}_\mu(t,\vec{r})]^k [\bar{\mathcal{X}}(t,\vec{r})\gamma^\mu T_k \mathcal{X}(t,\vec{r})] \,. \tag{C13}$$

# Appendix D: Massive vector field in SAQFT

In this Appendix, we give a brief description of the massive vector field in SAQFT. The particle associated with this quantum field (e.g., vector meson) has three degrees of freedom, which we designate as $\vec{V}_\mu(t,\vec{r}) = \left\{V_\mu^0, V_\mu^\pm\right\}$. It is constructed using the solution space of the free wave equation: $\left(\vec{\nabla}^2 - \partial_t^2\right)\vec{V}_\mu(t,\vec{r}) = M^2 \vec{V}_\mu(t,\vec{r})$. The space is reduced from four degrees of freedom carried by each of the components $\left\{\vec{V}_\mu\right\}$ to three by imposing the Lorenz invariant divergence condition $\sum_{\mu=0}^{\mu=3} \partial^\mu \vec{V}_\mu = 0$.

The quantum field associated with the massive vector bosons that possess structure is written as follows

$$\vec{V}_\mu(t,\vec{r}) = \int_\Omega e^{-\mathrm{i}Et}\vec{\psi}_\mu(E,\vec{r})\,\vec{a}(E)\,dE + \sum_{j=0}^{N} e^{-\mathrm{i}E_j t}\vec{\psi}_\mu^{\,j}(\vec{r})\,\vec{a}_j\,, \tag{D1}$$

where $\Omega = \{E : |E| \geq M\}$ and $\left\{|E_j| < M\right\}_{j=0}^{N}$. The creation and annihilation operators satisfy the following commutation relations

$$\left[a_s(E), a_{s'}^\dagger(E')\right] = a_s(E)\,a_{s'}^\dagger(E') - a_s^\dagger(E')\,a_{s'}(E) = \delta_{s,s'}\,\delta(E-E')\,,\ \left[a_i^s, (a_j^{s'})^\dagger\right] = \lambda^{-1}\delta_{s,s'}\,\delta_{i,j}\,. \tag{D2}$$

where $s$ and $s'$ stand for the three-vector indices $0,+,-$. All other commutators vanish. The continuous and discrete Fourier kernels are written as follows

$$\psi_\mu^s(E,\vec{r}) = \sum_{n=0}^{\infty} f_n^s(E)\,\phi_n^\mu(\vec{r}) = f_0^s(E)\sum_{n=0}^{\infty} p_n^s(z)\,\phi_n^\mu(\vec{r})\,, \tag{D3a}$$

$$\psi_\mu^{s,j}(\vec{r}) = \sum_{n=0}^{\infty} g_n^s(E_j)\,\phi_n^\mu(\vec{r}) = g_0^s(E_j)\sum_{n=0}^{\infty} p_n^s(z_j)\,\phi_n^\mu(\vec{r})\,, \tag{D3b}$$

where the spectral parameter $z$ is to be determined and $\left\{f_n^s(E), g_n^s(E_j)\right\}$ are real. The free wave equation $\left(\partial_t^2 - \vec{\nabla}^2 + M^2\right)V_\mu^s(t,\vec{r}) = 0$ becomes a three-term recursion relation for the spectral polynomials $\left\{p_n^s(z)\right\}$ dictating that

$$-\vec{\nabla}^2\phi_n^\mu(\vec{r}) = \alpha_n^s\phi_n^\mu(\vec{r}) + \beta_{n-1}^s\phi_{n-1}^\mu(\vec{r}) + \beta_n^s\phi_{n+1}^\mu(\vec{r})\,. \tag{D4}$$

The divergence condition, $\partial^\mu \vec{V}_\mu = 0$, becomes $E\phi_n^0(\vec{r}) = -\mathrm{i}\vec{\nabla}\cdot\vec{\phi}_n(\vec{r})$ and we require that:

For $E > -M$: $$-\mathrm{i}\vec{\nabla}\cdot\vec{\phi}_n(\vec{r}) = \left(c_n - M\right)\phi_n^0(\vec{r}) + d_n\phi_{n-1}^0(\vec{r})\,, \tag{D5a}$$

For $E < +M$: $$-\mathrm{i}\vec{\nabla}\cdot\vec{\phi}_n(\vec{r}) = \left(c_n + M\right)\phi_n^0(\vec{r}) + d_{n+1}\phi_{n+1}^0(\vec{r})\,, \tag{D5b}$$

where $\{c_n, d_n\}$ are non-zero constant parameters. Substituting (D5) in the divergence condition, we obtain:

For $E > -M$: $$\left(E + M\right)p_n^s(z) = c_n\,p_n^s(z) + d_{n+1}\,p_{n+1}^s(z)\,. \tag{D6a}$$

For $E < +M$: $$\left(E - M\right)p_n^s(z) = c_n\,p_n^s(z) + d_n\,p_{n-1}^s(z)\,. \tag{D6b}$$

Therefore, we end up with two possible spectral polynomials that satisfy either one of the following two recursion relations:

$$\left(E^2 - M^2\right)q_n^1(z) = \left(c_n^2 + d_n^2\right)q_n^1(z) + \left(c_{n-1}\,d_n\right)q_{n-1}^1(z) + \left(c_n\,d_{n+1}\right)q_{n+1}^1(z)\,. \tag{D7a}$$

$$\left(E^2 - M^2\right)q_n^2(z) = \left(c_n^2 + d_{n+1}^2\right)q_n^2(z) + \left(c_n\,d_n\right)q_{n-1}^2(z) + \left(c_{n+1}\,d_{n+1}\right)q_{n+1}^2(z)\,. \tag{D7b}$$

These two recursion relations must be associated with two out of the three spectral polynomials $p_n^{\pm}(z)$ and $p_n^0(z)$. So, now the questions are: First, which two out of these three polynomials correspond to (D7a) and (D7b)? Second, how to obtain the third recursion relation? The answer to the first question is easier since in the limit of zero mass, we should recover the two components of the massless vector field in Appendix C. That is, as $M \to 0$ we obtain $\vec{V}_\mu(t,\vec{r}) \to V_\mu^{\pm}(t,\vec{r}) = A_\mu^{\rightleftarrows}(t,\vec{r})$. Therefore, we must conclude that the spectral polynomials $p_n^+(z) = q_n^1(z)$ and $p_n^-(z) = q_n^2(z)$ with $z = E^2 - M^2$. To answer the second question, we resort to the hierarchy structure of supersymmetric tridiagonal systems associated with the spectral polynomials $q_n^1(z)$ and $q_n^2(z)$. Recently, Yamani and Mouayn carried out an elegant investigation of such a hierarchy structure in [37] according to which, we can write (see Table 2 therein)

$$\left(E^2 - M^2\right) p_n^0(z) = \left(c_{n+1}^2 + d_n^2\right) p_n^0(z) + \left(c_n\, d_n\right) p_{n-1}^0(z) + \left(c_{n+1}\, d_{n+1}\right) p_{n+1}^0(z). \qquad \text{(D7c)}$$

It is interesting to note that the recursion coefficient in relations (D7b) and (D7c) are obtained from each other by the exchange $c_n \leftrightarrow d_n$ making $\beta_n^0 = \beta_n^- = c_{n+1}\, d_{n+1}$ but $\alpha_n^0 \neq \alpha_n^-$.

The conjugate quantum field, the associated particles and anti-particles, and the Feynman propagators for these massive vector bosons are constructed in a manner analogous to that of Appendices C.